\documentclass[twocolumn,twocolappendix]{aastex631}

\usepackage{showyourwork}

\let\tablenum\relax             
\usepackage{siunitx}
\usepackage{blindtext}          
\usepackage{xcolor}
\usepackage[T1]{fontenc}

\usepackage[toc,nogroupskip,nohypertypes={glossary}]{glossaries} 
\glsdisablehyper

\newglossaryentry{HST}{name={HST}, description={Hubble Space Telescope}, first={Hubble Space Telescope (HST)}}
\newglossaryentry{NIR}{name={NIR}, description={Near-Infrared}, first={near-infrared (NIR)}}
\newglossaryentry{MIR}{name={MIR}, description={Mid-Infrared}, first={mid-infrared (MIR)}}
\newglossaryentry{FIR}{name={FIR}, description={Far-Infrared}, first={far-infrared (FIR)}}
\newglossaryentry{UV}{name={UV}, description={Ultraviolet}, first={ultraviolet (UV)}}
\newglossaryentry{FUV}{name={FUV}, description={Far-Ultraviolet}, first={far-ultraviolet (FUV)}}
\newglossaryentry{NUV}{name={NUV}, description={Near-Ultraviolet}, first={near-ultraviolet (NUV)}}
\newglossaryentry{EUV}{name={EUV}, description={extreme-Ultraviolet}, first={extreme-ultraviolet (EUV)}}
\newglossaryentry{SED}{name={SED}, description={Spectral energy distribution}, first={spectral energy distribution (SED)}, firstplural={spectral energy distributions (SEDs)}}
\newglossaryentry{ppm}{name={ppm}, description={parts per million}, first={parts per million (ppm)}}
\newglossaryentry{PDF}{name={PDF}, description={Probability density function}, first={probability density function (PDF)}, firstplural={probability density functions (PDF)}}
\newglossaryentry{PLATO}{name={PLATO}, description={PLAnetary Transits and Oscillations of stars}, first={PLATO~\citep[][]{Rauer2014}}}
\newglossaryentry{PSF}{name={PSF},description={Point Spread Function}, first={point spread function (PSF)}, firstplural={point spread functions (PSF)}}
\newglossaryentry{SFD}{name={SFD}, description={Size frequency distribution}, first={size frequency distribution (SFD)}, firstplural={Size frequency distributions (SFD)}}
\newglossaryentry{SNR}{name={SNR}, description={Signal-to-noise ratio}, first={signal-to-noise ratio (SNR)}}
\newglossaryentry{RV}{name={RV}, description={Radial Velocity}, first={Radial Velocity (RV)}, firstplural={Radial Velocities (RV)}}
\newglossaryentry{TTV}{name={TTV}, description={Transit Timing Variation}, first={transit timing variation (TTV)}, firstplural={transit timing variations (TTVs)}}
\newglossaryentry{TESS}{name={TESS}, description={Transiting Exoplanet Survey Satellite}, first={Transiting Exoplanet Survey Satellite~\citep[\textit{TESS},][]{Ricker2014}}}
\newglossaryentry{TIC}{name={TIC}, description={\textit{TESS} Input Catalog}, first={\textit{TESS} Input Catalog~\citep[TIC,][]{Stassun2018}}}
\newglossaryentry{JWST}{name={JWST}, description={James Webb Space Telescope}, first={James Webb Space Telescope~\citep[\textit{JWST},][]{Beichman2014}}}
\newglossaryentry{HWO}{name={HWO}, description={Habitable Worlds Observatory}, first={Habitable Worlds Observatory (HWO)}}
\newglossaryentry{HZ}{name={HZ}, description={habitable zone}, first={habitable zone (HZ)}}
\newglossaryentry{EEC}{name={EEC}, description={exo-Earth candidate}, first={exo-Earth candidate (EEC)}, firstplural={exo-Earth candidates (EEC)}}
\newglossaryentry{ool}{name={OoL}, description={Origins of Life}, first={Origins of Life (OoL)}}

\usepackage{datatool, filecontents}
\DTLsetseparator{,}

\DTLloaddb[noheader, keys={thekey,thevalue}]{variables}{variables.dat}
\newcommand{\var}[1]{\ensuremath{\DTLfetch{variables}{thekey}{#1}{thevalue}}}

\usepackage{comment}						 
\includecomment{comment}
\specialcomment{note}{\begingroup\sffamily\color{gray}}{\endgroup}

\usepackage[backgroundcolor=red!20!green!40!blue!10, textsize=tiny]{todonotes}
\usepackage{regexpatch}
\makeatletter
\xpatchcmd{\@todo}{\setkeys{todonotes}{#1}}{\setkeys{todonotes}{inline,#1}}{}{}
\makeatother

\newcommand{\rev}[1]{{#1}}              

\newcommand{\revv}[1]{{#1}}            

\DeclareSIUnit\mSun{M_\odot}
\DeclareSIUnit\Msun{M_\odot}
\DeclareSIUnit\mStar{M_\star}
\DeclareSIUnit\Mstar{M_\star}
\DeclareSIUnit\mEarth{M_\oplus}
\DeclareSIUnit\Mearth{M_\oplus}
\DeclareSIUnit\rEarth{R_\oplus}
\DeclareSIUnit\Rearth{R_\oplus}
\DeclareSIUnit\year{yr}
\DeclareSIUnit\au{au}
\DeclareSIUnit\dex{dex}
\DeclareSIUnit\ppm{ppm}
\DeclareSIUnit\eV{eV}
\DeclareSIUnit\parsec{pc}
\DeclareSIUnit\photons{photons}
\DeclareSIUnit\erg{erg}

\newcommand{\code}[1]{\texttt{#1}}
\newcommand{\project}[1]{\textsl{#1}}

\newcommand{\bioverse}{\code{Bioverse}}

\newcommand{\life}{\project{LIFE}}

\newcommand{\gaia}{\project{Gaia}}

\begin{document}
\begin{filecontents*}{variables.dat}
N_nautilus,214
uv_inhabited_M,148
semian_Nsamp1,10
semian_Nsamp2,100
uv_inhabited_FGK,99
d_max,270
M_G_max,16
M_st_max,1.5
f_life,0.8
NUV_thresh,300.0
deltaT_min,1
uv_inhabited,80
dlnZ_M,5.86
dlnZ_FGK,1.73
selectivity_FGK,-0.47
selectivity_M,-0.02
dlnZ_all,-0.4
p_FGK,0.0
p_M,0.0
nautilus_max_nuv,5\%
nautilus_S,5\%
nautilus_T_eff_st,50.0
p_all,0.0
p_FGK_dT100,0.5
uv_inhabited_FGK_dT100,0
d_max_dT100,270
M_G_max_dT100,16
M_st_max_dT100,1.5
f_life_dT100,0.8
NUV_thresh_dT100,300.0
deltaT_min_dT100,100
semian_Nsamp1_dT100,10
semian_Nsamp2_dT100,100
p_M_dT100,0.0
uv_inhabited_M_dT100,126
p_FGK_dT1000,0.5
uv_inhabited_FGK_dT1000,0
d_max_dT1000,270
M_G_max_dT1000,16
M_st_max_dT1000,1.5
f_life_dT1000,0.8
NUV_thresh_dT1000,300.0
deltaT_min_dT1000,1000
semian_Nsamp1_dT1000,10
semian_Nsamp2_dT1000,100
p_M_dT1000,0.5
uv_inhabited_M_dT1000,0
N_nautilus{file_suffix},214
N_nautilus_dT100,214
uv_inhabited_all,145
rimmer_threshold,43
rimmer_uncertainty,23
photon_flux_ratio,2.55
integrated_flux_ratio,2.45
stellar_sample_size_FGK,666
N_M_FGK,10123
f_M_FGK,0.89
median_M_st_M_FGK,0.17
p16_M_st_M_FGK,0.07
p84_M_st_M_FGK,0.18
median_T_eff_st_M_FGK,3055
p16_T_eff_st_M_FGK,346
p84_T_eff_st_M_FGK,337
N_L_FGK,551
f_L_FGK,0.05
median_M_st_L_FGK,0.07
p16_M_st_L_FGK,0.0
p84_M_st_L_FGK,0.0
median_T_eff_st_L_FGK,2006
p16_T_eff_st_L_FGK,46
p84_T_eff_st_L_FGK,268
N_K_FGK,666
f_K_FGK,0.06
median_M_st_K_FGK,0.63
p16_M_st_K_FGK,0.05
p84_M_st_K_FGK,0.05
median_T_eff_st_K_FGK,4068
p16_T_eff_st_K_FGK,88
p84_T_eff_st_K_FGK,336
stellar_sample_size_M,10123
N_M_M,10123
f_M_M,0.89
median_M_st_M_M,0.17
p16_M_st_M_M,0.07
p84_M_st_M_M,0.18
median_T_eff_st_M_M,3055
p16_T_eff_st_M_M,346
p84_T_eff_st_M_M,337
N_L_M,551
f_L_M,0.05
median_M_st_L_M,0.07
p16_M_st_L_M,0.0
p84_M_st_L_M,0.0
median_T_eff_st_L_M,2006
p16_T_eff_st_L_M,46
p84_T_eff_st_L_M,268
N_K_M,666
f_K_M,0.06
median_M_st_K_M,0.63
p16_M_st_K_M,0.05
p84_M_st_K_M,0.05
median_T_eff_st_K_M,4068
p16_T_eff_st_K_M,88
p84_T_eff_st_K_M,336
N_M_FGK_dT100,10123
f_M_FGK_dT100,0.89
median_M_st_M_FGK_dT100,0.17
p16_M_st_M_FGK_dT100,0.07
p84_M_st_M_FGK_dT100,0.18
median_T_eff_st_M_FGK_dT100,3055
p16_T_eff_st_M_FGK_dT100,346
p84_T_eff_st_M_FGK_dT100,337
N_L_FGK_dT100,551
f_L_FGK_dT100,0.05
median_M_st_L_FGK_dT100,0.07
p16_M_st_L_FGK_dT100,0.0
p84_M_st_L_FGK_dT100,0.0
median_T_eff_st_L_FGK_dT100,2006
p16_T_eff_st_L_FGK_dT100,46
p84_T_eff_st_L_FGK_dT100,268
N_K_FGK_dT100,666
f_K_FGK_dT100,0.06
median_M_st_K_FGK_dT100,0.63
p16_M_st_K_FGK_dT100,0.05
p84_M_st_K_FGK_dT100,0.05
median_T_eff_st_K_FGK_dT100,4068
p16_T_eff_st_K_FGK_dT100,88
p84_T_eff_st_K_FGK_dT100,336
N_M_M_dT100,10123
f_M_M_dT100,0.89
median_M_st_M_M_dT100,0.17
p16_M_st_M_M_dT100,0.07
p84_M_st_M_M_dT100,0.18
median_T_eff_st_M_M_dT100,3055
p16_T_eff_st_M_M_dT100,346
p84_T_eff_st_M_M_dT100,337
N_L_M_dT100,551
f_L_M_dT100,0.05
median_M_st_L_M_dT100,0.07
p16_M_st_L_M_dT100,0.0
p84_M_st_L_M_dT100,0.0
median_T_eff_st_L_M_dT100,2006
p16_T_eff_st_L_M_dT100,46
p84_T_eff_st_L_M_dT100,268
N_K_M_dT100,666
f_K_M_dT100,0.06
median_M_st_K_M_dT100,0.63
p16_M_st_K_M_dT100,0.05
p84_M_st_K_M_dT100,0.05
median_T_eff_st_K_M_dT100,4068
p16_T_eff_st_K_M_dT100,88
p84_T_eff_st_K_M_dT100,336
\end{filecontents*}

\title{Bioverse: Potentially Observable Exoplanet Biosignature Patterns Under the UV Threshold Hypothesis for the Origin of Life}

\author[0000-0001-8355-2107]{Martin Schlecker}
\affiliation{Steward Observatory, The University of Arizona, Tucson, AZ 85721, USA; \href{mailto:schlecker@arizona.edu}{schlecker@arizona.edu}}
\author[0000-0003-3714-5855]{D\'{a}niel Apai}
\affiliation{Steward Observatory, The University of Arizona, Tucson, AZ 85721, USA; \href{mailto:schlecker@arizona.edu}{schlecker@arizona.edu}}
\affiliation{Lunar and Planetary Laboratory, The University of Arizona, Tucson, AZ 85721, USA}
\author[0000-0003-3481-0952]{Antonin Affholder}
\affiliation{Department of Ecology and Evolutionary Biology, University of Arizona, Tucson AZ, USA}
\author[0000-0002-5147-9053]{Sukrit Ranjan}
\affiliation{Lunar and Planetary Laboratory, The University of Arizona, Tucson, AZ 85721, USA}
\affiliation{Blue Marble Space Institute of Science, Seattle, 98104, USA}
\author[0000-0002-5806-5566]{R\'{e}gis Ferri\`{e}re}
\affiliation{Department of Ecology and Evolutionary Biology, University of Arizona, Tucson AZ, USA}
\affiliation{Institut de Biologie de l'École Normale Supérieure, ENS, PSL, Paris, France}
\affiliation{International Research Laboratory for Interdisciplinary Global Environmental Studies (iGLOBES), CNRS, ENS, PSL, University of Arizona, Tucson AZ, USA}
\author[0000-0003-3702-0382]{Kevin K.\ Hardegree-Ullman}
\affiliation{Steward Observatory, The University of Arizona, Tucson, AZ 85721, USA; \href{mailto:schlecker@arizona.edu}{schlecker@arizona.edu}}
\affiliation{Caltech/IPAC-NASA Exoplanet Science Institute, 1200 E. California Blvd., MC 100-22, Pasadena, CA 91125, USA}
\author[0000-0002-3286-7683]{Tim Lichtenberg}
\affiliation{Kapteyn Astronomical Institute, University of Groningen, PO Box 800, 9700 AV Groningen, The Netherlands}
\author[0000-0003-3557-6256]{St\'{e}phane Mazevet}
\affiliation{Observatoire de la C\^{o}te d'Azur, Universit\'{e} C\^{o}te d'Azur, Nice, France}

\begin{abstract}
A wide variety of scenarios for the origin of life have been proposed, with many influencing the prevalence and distribution of biosignatures across exoplanet populations.
This relationship suggests these scenarios can be tested by predicting biosignature distributions and comparing them with empirical data.
Here, we demonstrate this approach by focusing on the cyanosulfidic origins-of-life scenario and investigating the hypothesis that a minimum near-ultraviolet (NUV) flux is necessary for abiogenesis.
Using Bayesian modeling and the \bioverse\ survey simulator, we constrain the probability of obtaining strong evidence for or against this ``UV Threshold Hypothesis'' with future biosignature surveys.

Our results indicate that a correlation between past NUV flux and current biosignature occurrence is testable for sample sizes of $\gtrsim$50 planets.
The diagnostic power of such tests is critically sensitive to the intrinsic abiogenesis rate and host star properties, particularly maximum past NUV fluxes.
Surveys targeting a wide range of fluxes, and planets orbiting M dwarfs enhance the chances of conclusive results, with sample sizes $\gtrsim$100 providing $\gtrsim$80\% likelihood of strong evidence if abiogenesis rates are high and the required NUV fluxes are moderate.
For required fluxes exceeding a few hundred erg/s/cm$^2$, both the fraction of inhabited planets and the diagnostic power sharply decrease.

Our findings demonstrate the potential of exoplanet surveys to test origins-of-life hypotheses.
 Beyond specific scenarios, this work underscores the broader value of realistic survey simulations for future observatories (e.g., HWO, LIFE, ELTs, Nautilus) in identifying testable science questions, optimizing mission strategies, and advancing theoretical and experimental studies of abiogenesis.
\end{abstract}

\section{Introduction}
\label{sec:intro}
\rev{The probability of abiogenesis remains one of the fundamental questions in modern science.
As of now, it has been primarily explored through statistical arguments~\citep[e.g.,][]{Spiegel2012,Kipping2021,Lingam2024}.
At the same time, a} wide variety of scenarios for the origin of life have been proposed~\citep[e.g.,][]{Baross1985,Brasier2011,Mulkidjanian2012,Fox2013,Deamer2015,Westall2018}.
While we may still be far from conclusively testing these scenarios, new prospects in the search for conditions favorable to life have opened up by thinking of the origin of life as a planetary phenomenon and identifying global-scale environmental properties that might support pathways to life~\citep{Sasselov2020}.
In particular, specific planetary conditions are needed to create stockpiles of initial compounds for prebiotic chemistry; and planetary processes are required to trigger the prebiotic synthesis.
Such planetary conditions can be hypothesized for exoplanets located in the \gls{HZ} of their host star, with persistent liquid water on their surface.
For example, if deep-sea or sedimentary hydrothermalism is required for abiogenesis, then the insulation of an ocean from the planetary crust minerals (e.g., due to high-pressure ices) may reduce or eliminate the chances of life emerging~\citep[e.g.,][]{Baross1985}.
The alternate scenario of a surface locally subject to wet-dry cycles requires a planetary exposure to mid-range \gls{UV} irradiation, as a source of energy and an agent of selection in chemical evolution~\citep[e.g.,][]{Deamer2019}.
This ``UV Threshold Hypothesis'' states that \gls{UV} light in a specific wavelength range played a constructive role in getting life started on Earth~\citep{Ranjan2016,Ranjan2017c,Rimmer2018,Rapf2016}, and it could provide a probabilistic approach to the interpretation of possible future biosignature detections~\citep[e.g.,][]{Catling2018,Walker2018}.

The association of chemical pathways to life and planetary environmental conditions offers a new opportunity to test alternate scenarios for life emergence based on planetary-level data collected from the upcoming observations of populations of exoplanets.
Deep-sea hydrothermal scenarios require planetary conditions that may not be met on ocean worlds with large amounts of water, where the water pressure on the ocean floor is high enough to form high-pressure ices~\citep{Noack2016,Kite2018}.  
In this case, a testable prediction would be that planets with high-pressure ices do not show biosignatures.
Likewise, if \gls{UV} light is required to get life started, then there is a minimum planetary \gls{UV} flux requirement to have an inhabited world.
This requirement is set by competing thermal processes; if the photoreaction does not move forward at a rate faster than the competitor thermal process(es), then the abiogenesis scenario cannot function.
On the other hand, abundant \gls{UV} light vastly in excess of this threshold does not increase the probability of abiogenesis, since once the \gls{UV} photochemistry is no longer limiting, some other thermal process in the reaction network will be the rate-limiting process instead.
Therefore, a putative dependence of life on \gls{UV} light is best described as a step function~\citep[e.g.,][]{Ranjan2017c,Rimmer2018,Rimmer2021}.

The goal of this work is to evaluate the potential of future exoplanet surveys to test the hypothesis that a minimum past \gls{NUV} flux is required for abiogenesis.
We focus on one version of the UV Threshold Hypothesis, the so-called cyanosulfidic scenario, which has been refined to the point where the required threshold flux has been measured \rev{to be $\SI{6.8\pm3.6 e10}{\photons\per\centi\meter\squared\per\second\per\nano\meter}$ integrated} from \SIrange{200}{280}{\nano\meter} at the surface~\citep{Rimmer2018,Rimmer2021a,Rimmer2023,Ranjan2023a}\rev{\footnote{or $\sim$\var{rimmer_threshold}$\pm$\var{rimmer_uncertainty}$\,$\SI{}{\erg\per\second\per\centi\meter\squared} in energy flux units.}}.
\rev{While prebiotic photochemistry is fundamentally driven by photon number flux, we choose to frame the problem in terms of energy flux, for which empirical estimates in the \gls{NUV} exist across stellar populations and ages~\citep{Richey-Yowell2023}.
Due to the inherent uncertainties concerning the relationship between surface and top-of-atmosphere fluxes, we treat the threshold \gls{NUV} flux as a free parameter.}

We \rev{first} follow a semi-analytical Bayesian analysis to estimate probabilities of obtaining strong evidence for or against \rev{the UV Threshold Hypothesis}.
Under \rev{this hypothesis} ($H_1$), the probability of an exoplanet having detectable biosignatures is zero if the \rev{\gls{NUV}} irradiation is less than the threshold, and it is equal to the (unknown) probability of \rev{life emerging and persisting}, $f_\mathrm{life}$, if \gls{NUV} exceeds the threshold for a sufficiently long period of time.
Under the null hypothesis ($H_\mathrm{null}$), that probability simply is $f_\mathrm{life}$, that is, it does not correlate with the \gls{UV} flux.

\begin{figure*}
    \begin{centering}
        \includegraphics[width=\hsize]{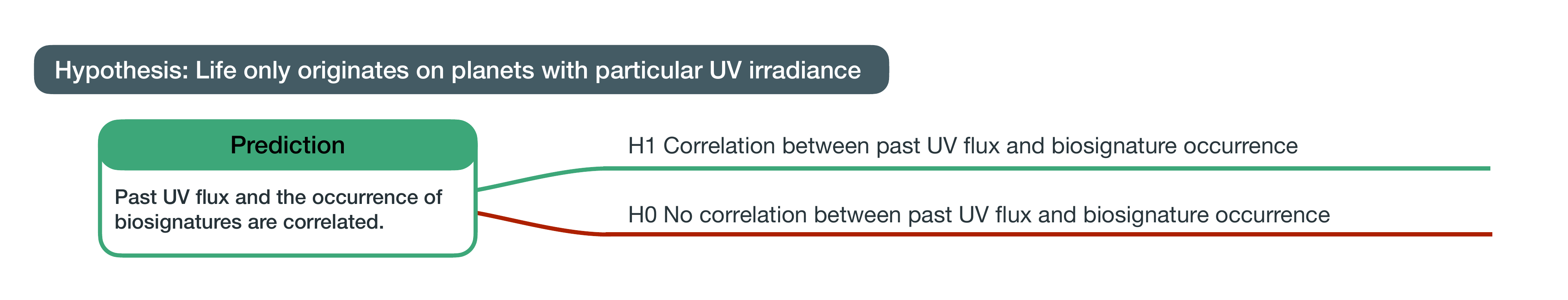}
        \caption{UV~Threshold Hypothesis and null hypothesis derived from the cyanosulfidic scenario.}
        \label{fig:hypotheses}
    \end{centering}
\end{figure*}
Figure~\ref{fig:hypotheses} shows these hypotheses as derived from the predictions of the cyanosulfidic scenario.
Given a sample of planets, where for some of them we have convincing biosignature detections but remain agnostic on $f_\mathrm{life}$, we ask what evidence for $H_1$ and $H_\mathrm{null}$ we can expect to obtain.

A real exoplanet survey will be subject to observational biases and sample selection effects, and will be constrained by the underlying demographics of the planet sample.
To assess the information gain of a realistic exoplanet survey, we employed \bioverse~\citep{Bixel2021,Hardegree-Ullman2023,Schlecker2024,Hardegree-Ullman2025}, a framework that integrates multiple components including statistically realistic simulations of exoplanet populations, a survey simulation module, and a hypothesis testing module to evaluate the statistical power of different observational strategies.

This paper is organized as follows:
In Section~\ref{sec:methods}, we introduce both our semi-analytical approach and \bioverse\ simulations for testing the UV Threshold Hypothesis.
Section~\ref{sec:results} presents the results of these experiments for a generic survey as well as for a realistic transit survey.
In Section~\ref{sec:discussion}, we discuss our findings before concluding with a summary in Section~\ref{sec:conclusions}.



\section{Methods}
\label{sec:methods}

\subsection{Fraction of inhabited planets with detectable biosignatures}
%
%
Here, we conduct a theoretical experiment on the UV~Threshold Hypothesis by relating the occurrence of life on an exo-earth candidate with a minimum past quiescent stellar \gls{UV} flux, focusing on the prebiotically interesting \gls{NUV} range from \SIrange{200}{280}{\nano\meter}~\citep{Ranjan2016}.
Our core hypothesis shall be that life only occurs on planets that at some point in their history have received such radiation at a flux exceeding a threshold $F_\mathrm{NUV, min}$.

\subsection{Semi-analytical approach}\label{sec:met-semianalytical}
We first assessed the expected probabilities of obtaining true negative or true positive evidence for the UV Threshold Hypothesis ($H_1$) above, as well as the probability for misleading or inconclusive evidence, under idealized conditions.
This serves as a first-order estimate of the information content of a survey, before we take into account the effects of exoplanet demographics, sample selection, and survey strategy.

Presumably, not all habitable worlds are inhabited and not all inhabited worlds develop detectable biosignatures.
The fraction of \glspl{EEC} that are both inhabited and exhibit detectable biosignatures at the time of observation is unknown \rev{and is represented by }\revv{ $f_\mathrm{life}$.}
\revv{This term captures the combined probability that life \textit{both} emerges on a planet and persists in a form that produces detectable biosignatures by the time of observation.
}
\rev{Due to our ignorance about its true value or even its order of magnitude, we draw $f_\mathrm{life}$ from a log-uniform prior probability distribution} \revv{over the range $[10^{-15}, 1]$~\citep[][but see Appendix~\ref{sec:appendix-flife-prior} for a discussion of the impact of different priors]{Balbi2023a}}.
Let us consider the probability to detect a biosignature $P(L)$, and let our observable be the inferred past \gls{NUV} flux of the planet $F_\mathrm{NUV}$.
Under Hypothesis $H_1$, there exists a special unknown value of $F_\mathrm{NUV}$, noted $F_\mathrm{NUV, min}$ such that

\begin{align}
    P(L|F_\mathrm{NUV},H_1) &=  f_\mathrm{life} \quad \text{if } F_\mathrm{NUV}>F_\mathrm{NUV, min}\\
    P(L|F_\mathrm{NUV},H_1) &=  0               \quad  \text{otherwise.}
\end{align}
\revv{Here, the step-function dependence on $F_\mathrm{NUV}$ applies only to the initial step of abiogenesis, consistent with the cyanosulfidic scenario in which UV irradiation is required for the synthesis of key prebiotic precursors.
We do not assume that the persistence of life or the presence of detectable biosignatures depends on UV flux.}

The corresponding null hypothesis $H_\mathrm{null}$ is that there exists no such special value of $F_\mathrm{NUV}$ and that
\begin{equation}
P(L|F_\mathrm{NUV},H_\mathrm{null}) = f_\mathrm{life}.
\end{equation}
In other words, $H_\mathrm{null}$ states that $P(L)$ is independent of $F_\mathrm{NUV}$.

\rev{Next, we determine the probability distribution of sample outcomes, or likelihood of each hypothesis.}
Let $Y=\sum_i^n L_i$ be the random variable counting the number of positive life detections in a sample of size $n$. Its probability mass function under the null hypothesis $H_\mathrm{null}$ is that of a binomial distribution:
\begin{equation}
    \label{eq:semian:likelihoodHnull}
P(Y=k|H_\mathrm{null}) = \binom{n}{k}f_\mathrm{life}^k(1-f_\mathrm{life})^{n-k}.
\end{equation}

Under $H_1$, $Y$ also follows a binomial distribution, however it is conditioned by $n_{\lambda} = n(\{F_\mathrm{NUV, i} \text{ if } F_\mathrm{NUV, i}>F_\mathrm{NUV, min}\}_{i \in [1,n]})$, the number of values of $F_\mathrm{NUV}$ in the experiment that exceed $F_\mathrm{NUV, min}$
\begin{equation}
\label{eq:semian:likelihoodH1}
P(Y=k|H_1) = \binom{n_{\lambda}}{k}f_\mathrm{life}^k(1-f_\mathrm{life})^{n_{\lambda}-k}.
\end{equation}

\rev{
Following a similar approach as in \cite{affholder2025interior}, we aim to quantify the information gain from our sampling procedure by computing the Bayes factors~\citep{Jeffreys1939}
}
\begin{equation}\label{eq:bayes_factor}
BF_{H_1,H_\mathrm{null}} = \frac{P(Y=k|H_1)}{P(Y=k|H_\mathrm{null})} = \frac{\binom{n_\lambda}{k}}{\binom{n}{k}}(1-f_\mathrm{life})^{n_{\lambda}-n}
\end{equation}
and
\begin{equation}
    \label{eq:bayes_factor2}
BF_{H_\mathrm{null},H_1} = \frac{P(Y=k|H_\mathrm{null})}{P(Y=k|H_1)} = \frac{\binom{n}{k}}{\binom{n_\lambda}{k}}(1-f_\mathrm{life})^{n-n_{\lambda}}.
\end{equation}
Given a sample of planets, where for some of them we have convincing biosignature detections but remaining agnostic on $f_\mathrm{life}$: What evidence for $H_\mathrm{1}$ and $H_\mathrm{null}$ can we expect to get?
The analytical expression for the Bayes factor of this inference problem (Equation~\ref{eq:bayes_factor}) is determined by the unknown variables $f_\mathrm{life}$ and $F_\mathrm{NUV, min}$, as well as by the summary statistic $Y$ (number of biosignature detections).
To compute the distribution of evidences, we repeatedly generated samples under $H_\mathrm{1}$ and $H_\mathrm{null}$ and computed the Bayes factors $BF_{H_1,H_\mathrm{null}}$ and $BF_{H_\mathrm{null}, H_1}$.
We then evaluated the fraction of Monte Carlo runs in which certain evidence thresholds~\citep{Jeffreys1939} were exceeded.

Under a more realistic scenario, the distribution of $n_{\lambda}$ depends on additional planetary properties and their evolution, as well as on observational biases and sample selection effects of the survey.
We will address these in the following section.

\subsection{Exoplanet survey simulations with \bioverse}
To assess the diagnostic power of realistic exoplanet surveys, we employed our survey simulator and hypothesis testing framework \bioverse~\citep{Bixel2021}.
The general approach is as follows:
\begin{enumerate}
\item \textbf{Exoplanet population synthesis:} We populate the Gaia Catalogue of Nearby Stars~\citep{Smart2021} with synthetic exoplanets whose orbital parameters and planetary properties reflect our current understanding of exoplanet demographics~\citep{Bergsten2022,Hardegree-Ullman2023}.
Here, we also inject the demographic trend in question - in this case we assign biosignatures according to $H_1$, i.e., to planets in the \gls{HZ} that have received \gls{NUV} fluxes above a certain threshold.
    \item \textbf{Survey simulation:} We simulate the detection and characterization of these exoplanets with a hypothetical survey, taking into account the survey's sensitivity, target selection, and observational biases.
To model the sensitivity of the information gain of a proposed mission to sample selection and survey strategy, we conduct survey simulations with \bioverse\ using different sample sizes and survey strategies.
    \item \textbf{Hypothesis testing:} We evaluate the likelihood that a given survey would detect a specified demographic trend in the exoplanet population and estimate the precision with which the survey could constrain the parameters of that trend.
    A common definition of the null hypothesis $H_\mathrm{null}$, which is also applied here, is that there is no relationship between the independent variable (here: maximum \gls{NUV} flux) and the dependent variable (here: biosignature occurrence).
    The alternative hypothesis $H_1$ proposes a specific relationship between the independent and dependent variables.
    \bioverse\ offers either Bayesian model comparison or non-parametric tests to evaluate the evidence for or against the null hypothesis.
\end{enumerate}

To determine the diagnostic capability of a given survey, \bioverse\ runs multiple iterations of the simulated survey and calculates the fraction of realizations that successfully reject the null hypothesis.
We used this metric, known as the statistical power, to quantify the potential information content of the survey, identify critical design trades, and find strategies that maximize the survey's scientific return.

\subsubsection{Simulated star and planet sample}
We generated two sets of synthetic exoplanet populations, one for FGK-type stars and one for M-type stars.
The stellar samples are drawn from the Gaia Catalogue of Nearby Stars~\citep{Smart2021} with a maximum Gaia magnitude of \var{M_G_max} and a maximum stellar mass of \var{M_st_max}~\SI{}{\Msun}.
We included stars out to a maximum distance $d_{\max}$ that depends on the required planet sample size.
\revv{In our FGK sample, we find that K~dwarfs dominate with effective temperatures of $\var{median_T_eff_st_K_FGK}^{+\var{p84_T_eff_st_K_FGK}}_{-\var{p16_T_eff_st_K_FGK}}$~K and masses of $\var{median_M_st_K_FGK}^{+\var{p84_M_st_K_FGK}}_{-\var{p16_M_st_K_FGK}}$~$M_\odot$ (median and 16th/84th percentiles).
In the M~dwarf sample, the stars have effective temperatures of $\var{median_T_eff_st_M_M}^{+\var{p84_T_eff_st_M_M}}_{-\var{p16_T_eff_st_M_M}}$~K and masses of $\var{median_M_st_M_M}^{+\var{p84_M_st_M_M}}_{-\var{p16_M_st_M_M}}$~$M_\odot$.}

Planets were generated and assigned to the synthetic stars following the occurrence rates and size/orbit distributions of \citet{Bergsten2022}.
Following \citet{Bixel2021}, we considered only transiting \glspl{EEC} with radii $0.8\, S^{0.25} < R < 1.4 $ that are within the \gls{HZ} (see Section~\ref{sec:met-hz}).
The lower limit was suggested as a minimum planet size to retain an atmosphere~\citep{Zahnle2017}.
For all survey simulations and hypothesis tests, we repeated the above in a Monte Carlo fashion to generate randomized ensembles of synthetic star and planet populations~\citep[][]{Bixel2021}.

\subsubsection{Habitable zone occupancy and UV flux}\label{sec:met-hz}
To construct a test \rev{of} the UV~Threshold Hypothesis, we required that life occurs only on \revv{\gls{HZ}} planets with sufficient past \gls{UV} irradiation exceeding the origins of life threshold $F_\mathrm{NUV, min}$.
Further, we required this flux to have lasted for a minimum duration $\Delta T_\mathrm{min}$ to allow for a sufficient ``origins timescale''~\citep{Rimmer2023}.
\revv{While there is no consensus on the minimum timescale required for abiogenesis~\citep[e.g.,][]{Lazcano1996}, we adopt  a nominal value of} \rev{$\Delta T_\mathrm{min} = \SI{1}{\mega\year}$.}
\rev{Under $H_1$, longer origins timescales have minimal impact on a generic transit survey but significantly decrease the fraction of inhabited planets around FGK stars, as explored in Appendix~\ref{sec:long-origins}.
}
\revv{
The ``clock'' for the required duration $\Delta T_\mathrm{min}$ starts once the planet resides within the habitable zone and receives NUV flux above the threshold $F_\mathrm{NUV, min}$.
}

All commonly investigated origins-of-life scenarios require water as a solvent;
we thus considered only rocky planets that may sustain liquid water on their surface, i.e., that occupy their host star's momentary \gls{HZ} during the above period, as well as at the epoch of observation.
The \gls{HZ} describes a region around a star where a planet with Earth's atmospheric composition and climate feedbacks can maintain liquid water on its surface~\citep[e.g.,][]{Ramirez2017,Ramirez2018,MolLous2022,Spinelli2023,Tuchow2023}.
Here, we adopted orbital distance estimates that define the \gls{HZ} as the region between the runaway greenhouse transition, where the stellar instellation cannot anymore be balanced through infrared cooling to space~\citep{Ingersoll1969}, and the maximum greenhouse limit, corresponding to the maximum distance at which surface temperatures allowing liquid water can be maintained through a CO$_2$ greenhouse effect~\citep{Kasting1991,Kasting1993,Underwood2003,Kopparapu2013,Kopparapu2014}.
\rev{The exact boundaries of the habitable zone are known to be sensitive to the star's luminosity, spectral type, the planet's mass, and the planet's atmospheric properties~\citep[e.g.,][]{Pierrehumbert2011,Ramirez2014,Ramirez2017,Ramirez2018b,Koll2019,Ramirez2018,Ramirez2020,Bonati2021,Chaverot2022,Turbet2023}.
Here, we adopted the commonly used parametrization in \citet{Kopparapu2014} to derive luminosity and planetary mass-dependent distance limits of the \gls{HZ} $a_\mathrm{inner}$ and $a_\mathrm{outer}$.
}

\begin{figure*}
    \script{hz_nuv_evo.py}
    \begin{centering}
        \includegraphics[width=\hsize]{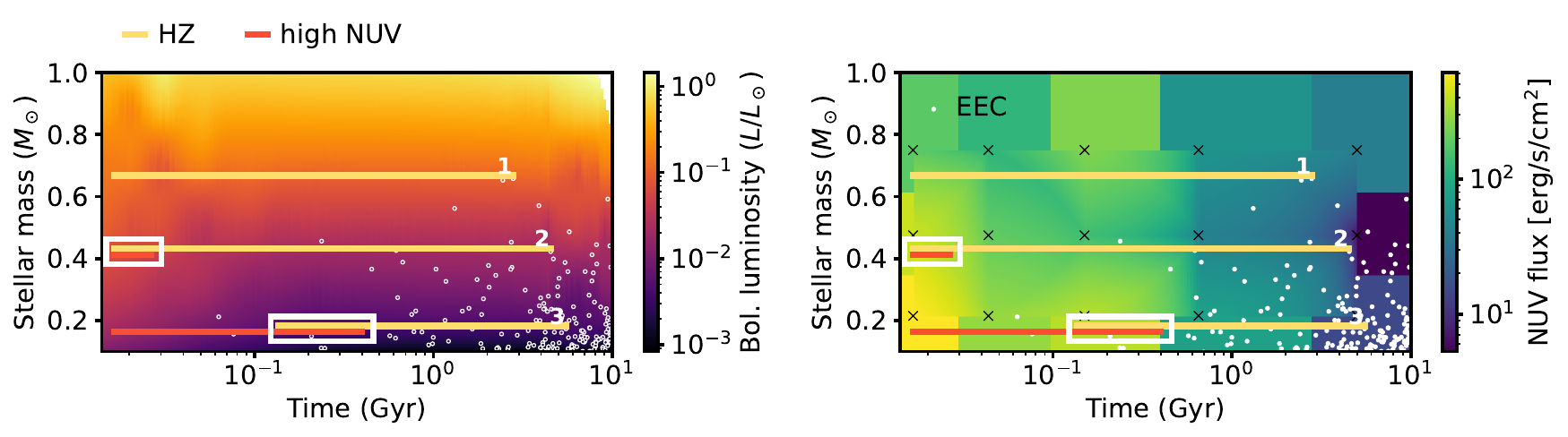}
        \caption{Interpolated stellar luminosity evolution (left) and evolution of the \gls{NUV} flux in the \gls{HZ} (right) as a function of host star mass.
        Scatter points show age and host star mass of the transiting planets in the synthetic planet sample; crosses denote the estimated \gls{NUV} values in \rev{\citet{Richey-Yowell2023}}. 
        We show three evolutionary tracks for a threshold flux of $F_\mathrm{NUV, min} = \var{NUV_thresh}\,\SI{}{\erg\per\second\per\centi\meter\squared}$ that occupy the \gls{HZ} (yellow sections) and exceed the threshold \gls{NUV} flux (red sections) at different times.
        Where these sections overlap (white rectangles), the requirements for abiogenesis are met and we assign a biosignature detection with probability $f_\mathrm{life}$.
        Planet~1 is an \gls{EEC} orbiting a K~dwarf that never receives sufficient \gls{NUV} flux for abiogenesis.
        Planet~2 and Planet~3 enter the \gls{HZ} at different times and receive sufficient \gls{NUV} flux for different durations until their respective host star evolves below the threshold.
        }
        \label{fig:hz_nuv_evo}
    \end{centering}
\end{figure*}

\revv{To determine \gls{HZ} occupancy, we took into account the evolution of the host star's luminosity and \gls{HZ} boundaries.
Specifically, we} interpolated the stellar luminosity evolution grid of \citet{Baraffe1998} using a Clough Tocher interpolant~\citep[][see left panel of Figure~\ref{fig:hz_nuv_evo}]{Nielson1983,Alfeld1984} to compute the evolution of the inner (runaway greenhouse) and outer (maximum greenhouse) edges as a function of planet mass and stellar spectral type~\citep{Kopparapu2014}.
Being a local interpolation method, Clough Tocher enables rapid processing while producing a smooth interpolating surface that highlights local trends.
From this, we get each planet's epochs within and outside the \gls{HZ}.

\revv{We note that in particular for late M dwarfs, the pre-main-sequence phase can reach into the age ranges considered here~\citep{Baraffe2015}. This may expose planets to elevated levels of radiation, potentially inducing runaway greenhouse effects, water loss, or atmospheric erosion~\citep{Ramirez2014,Luger2015}.
As such, the actual window of surface habitability may be shorter than the total time a planet spends in the habitable zone during its main-sequence lifetime.
We defer a detailed treatment of pre-main-sequence effects to future work.
}

For the \gls{NUV} flux, we used the age- and stellar mass-dependent \gls{NUV} fluxes in the \gls{HZ} obtained by \citet{Richey-Yowell2023}, which considers GALEX \gls{UV} data in the wavelength range of \SIrange{177}{283}{\nano\meter}.
We linearly interpolate in their measured grid, where we convert spectral type to stellar mass using the midpoints of their mass ranges (\SI{0.75}{\Msun} for K~stars, \SI{0.475}{\Msun} for early-type M~stars,  and \SI{0.215}{\Msun} for late-type M~stars).
Outside the age and stellar mass range covered in \citet{Richey-Yowell2023}, we extrapolate using nearest simplex (see right panel of Figure~\ref{fig:hz_nuv_evo}).

We then determined which planets were both in the \gls{HZ} and had \gls{NUV} fluxes above $F_\mathrm{NUV, min}$.
To avoid considering short transitional phases, we require this situation to last for a minimum duration $\Delta T_\mathrm{min} \geq \var{deltaT_min}\,\SI{}{\mega\year}$.
We assigned the emergence \rev{and persistence} of life to a random fraction $f_\mathrm{life}$ of all temperate planets fulfilling these requirements.
For the probability of a planet having detectable biosignatures, $P(\mathrm{bio})$, the UV~Threshold Hypothesis then states
\begin{equation}\label{eq:hypothesis-uv}
    H_{1} : P(\mathrm{bio}) =
        \begin{cases}
            0 , & F_\mathrm{NUV} < F_\mathrm{NUV, min}\\
            f_\mathrm{life}, & F_\mathrm{NUV} \geq F_\mathrm{NUV, min} \\
            & \text{and in HZ for} \, \Delta t \geq \var{deltaT_min} \,\SI{}{\mega\year}
        \end{cases}
\end{equation}
and the corresponding null hypothesis
$H_\mathrm{null} : P(\mathrm{bio}) = f_\mathrm{life},$
i.e., no correlation with \gls{UV} flux.

\subsubsection{Transit survey simulations}
With the synthetic star and planet samples generated, we used \bioverse's survey module to simulate noisy measurements of key observables with a transit survey.
We assumed a hypothetical mission that can target a large planet sample with high photometric precision and conduct a biosignature search on these planets~\citep[e.g.,][]{Apai2019,Apai2022}.
The simulated survey was designed to measure planetary instellation (for \gls{HZ} occupancy) with a precision of \var{nautilus_S} and host star effective temperature with a precision of \var{nautilus_T_eff_st}~K.
The maximum past \gls{NUV} flux a planet received can be determined within a precision of \var{nautilus_max_nuv}.
To marginalize over choices of biosignatures and their detectability, which are beyond the scope of this study, we assumed that any inhabited planet would show a biosignature detectable by the survey.


\subsubsection{Hypothesis testing}
\rev{We ought to choose a statistical test that is sensitive to the UV~Threshold Hypothesis, and that could be realistically conducted in a future transit survey (which may include auxiliary information from ground-based observations, archived data, or models).
Given the available types of data expected from such surveys, our test shall be non-parametric and compare two samples -- planets with and without biosignatures -- to assess whether they are drawn from the same underlying population in terms of their infered historic maximum \gls{NUV} flux.
Common options include the Kolmogorov-Smirnov test, the Brunner-Munzel test~\citep{Brunner2000}, and the Mann-Whitney U test~\citep{Mann1947}.
Due to its availability and suitability for large sample sizes, we chose the Mann-Whitney U test, which evaluates if one sample is stochastically greater than the other.
Here, we }
compare the distributions of \gls{NUV} fluxes of planets with and without biosignatures.
The implementation in \bioverse\ relies on the \texttt{scipy.stats.mannwhitneyu} function~\citep{Virtanen2020} and returns a p-value \rev{for each test.
To balance the trade-off between Type I and Type II error risks, we set the significance level to the widely adopted threshold $\alpha = 0.05$.
To quantify the diagnostic power of the survey, we conducted repeated randomized realizations and calculated the fraction of successful rejections of the null hypothesis, i.e., the statistical power.
}

\section{Results}
\label{sec:results}


\subsection{Semi-analytical assessment}\label{sec:results-semianalytical}
\begin{figure*}
    \script{semian_true_evidence.py}
    \begin{centering}
        \includegraphics[width=\hsize]{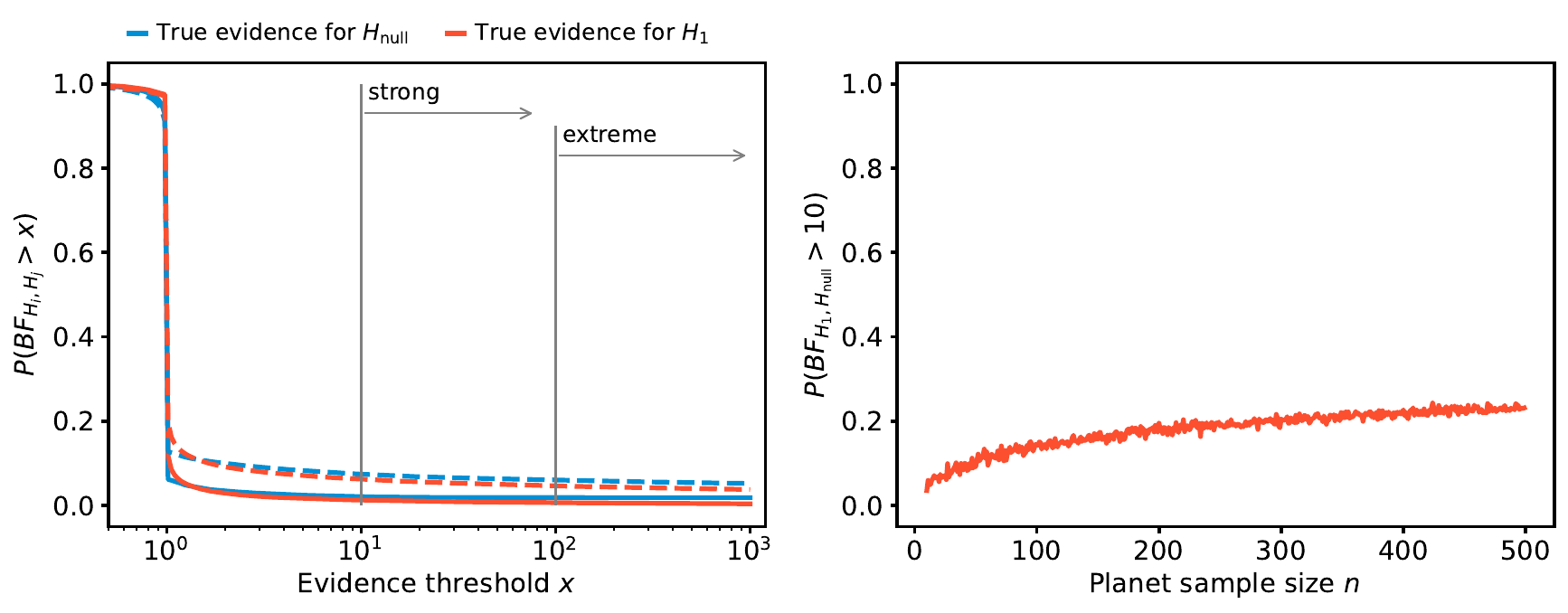}
        \caption{Obtaining true strong evidence with different sample sizes. Left: Probability to reach given evidence levels for $H_\mathrm{1}$ and $H_\mathrm{null}$ under sample sizes $n = \var{semian_Nsamp1}$ (solid) and $n = \var{semian_Nsamp2}$ (dashed). Vertical lines denote thresholds for ``strong'' evidence, $BF_{H_i, H_j}$ > 10, and ``extreme'' evidence, $BF_{H_i, H_j}$ > 100. Right: Probability of obtaining true strong evidence for $H_\mathrm{1}$ as a function of sample size $n$.}
        \label{fig:semian_true_evidence}
    \end{centering}
\end{figure*}

In Section~\ref{sec:met-semianalytical}, we computed the probability for true positive evidence for $H_\mathrm{1}$ and $H_\mathrm{null}$, respectively (Equations~\ref{eq:bayes_factor},~\ref{eq:bayes_factor2}).
Figure~\ref{fig:semian_true_evidence} shows how these evidences are distributed for sample sizes \var{semian_Nsamp1} and \var{semian_Nsamp2}, and how likely we are to obtain strong evidence ($BF_{H_i, H_j}$ > 10) \rev{in the agnostic case where we draw $f_\mathrm{life}$ from a log-uniform distribution \revv{over the range $[10^{-15}, 1]$}}.
For $n = \var{semian_Nsamp1}$, strong true evidence for $H_\mathrm{1}$ ($H_\mathrm{null}$) can be \rev{expected in $\sim \SI{3}{\percent}$ ($\sim \SI{6}{\percent}$) of} all random experiments.
In the majority of cases, the outcome of the survey will be inconclusive.
The situation improves with larger samples: \rev{for $n = \var{semian_Nsamp2}$, \SI{14}{\percent} (\SI{16}{\percent}) of} random samples permit conclusive inference (strong true evidence) \rev{under $H_\mathrm{1}$ ($H_\mathrm{null}$).}

The expected resulting evidence further depends on the a priori unknown \rev{rate of life's emergence and persistence} $f_\mathrm{life}$ and on the \gls{NUV} flux threshold.
\begin{figure*}
    \script{semian_true_evidence.py}
    \begin{centering}
        \includegraphics[width=\hsize]{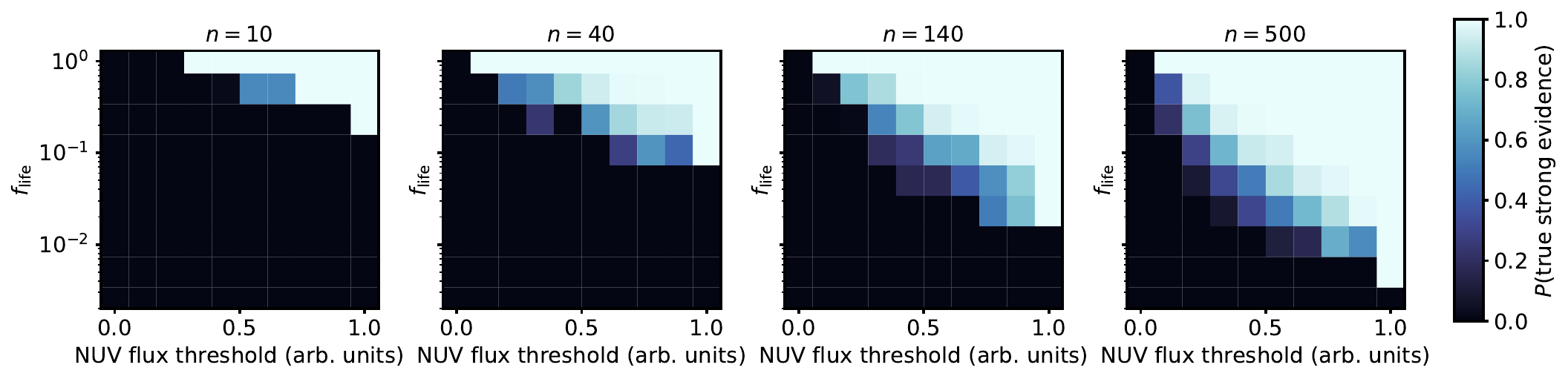}
        \caption{Probability of obtaining true strong evidence for different abiogenesis rates, \gls{NUV} flux thresholds, and sample sizes. For each of these parameters, higher values increase the probability of yielding strong evidence.}
        \label{fig:semian_evidence-grid}
    \end{centering}
\end{figure*}
Figure~\ref{fig:semian_evidence-grid} illustrates this dependency: For very low values of either parameter, samples drawn under the null or alternative hypotheses are indistinguishable and the Bayesian evidence is always low.
Both higher $f_\mathrm{life}$ and higher \gls{NUV} flux thresholds increase the probability of obtaining strong evidence.
Larger sample sizes enable this at lower values of these parameters.

\begin{figure*}
    \script{semian_true_evidence.py}
    \begin{centering}
        \includegraphics[width=\hsize]{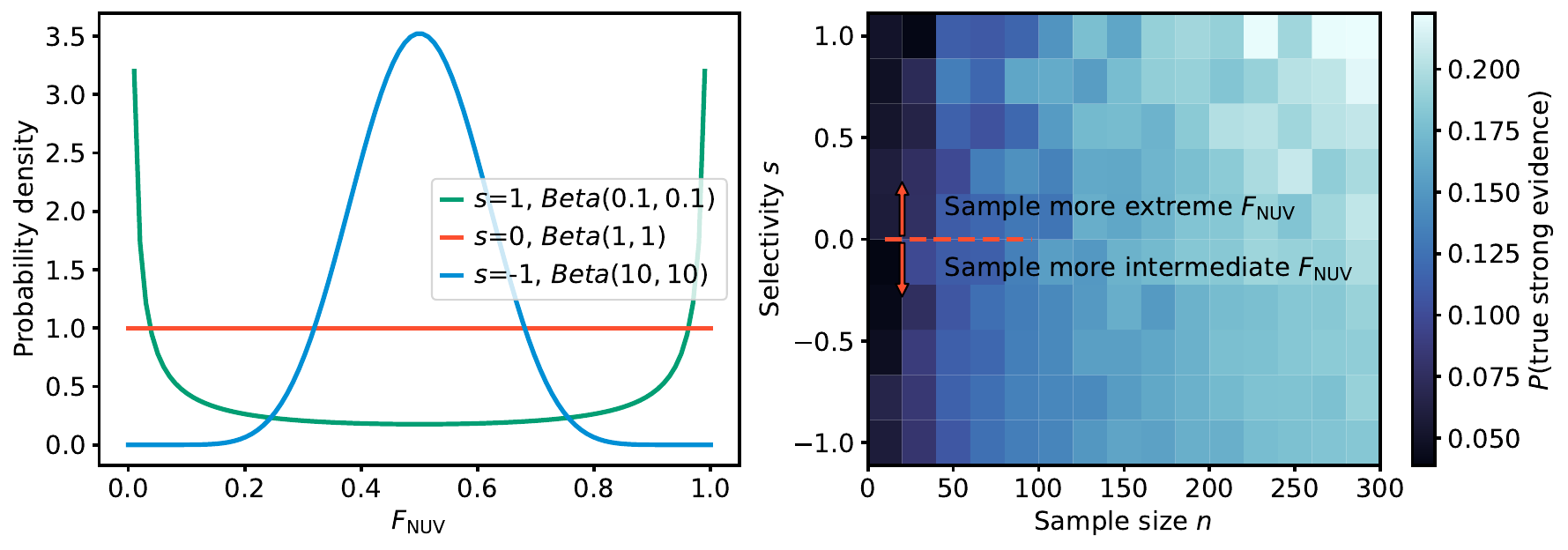}
        \caption{Scaling of the probability of obtaining true strong evidence with sample selectivity. Left: Sampling distribution for different selectivity parameters $s$. Right: Resulting \mbox{P(true strong evidence)}, where $f_\mathrm{life}$ and $F_\mathrm{NUV, min}$ are randomly drawn from \rev{log-uniform and uniform distributions, respectively}. Sampling more extreme values of $F_\mathrm{NUV}$ is more likely to yield strong evidence.}
        \label{fig:semian_selectivity}
    \end{centering}
\end{figure*}

So far, we have \rev{drawn random values from uniform distributions for} $F_\mathrm{NUV, min}$, and $F_\mathrm{NUV}$\rev{, and from a log-uniform distribution for $f_\mathrm{life}$}. 
\revv{Higher $f_\mathrm{life}$ increase} the evidence (see Equation~\ref{eq:bayes_factor}) but a survey strategy cannot influence \revv{this parameter}.
The same is true for $F_\mathrm{NUV, min}$, where again higher values increase the evidence as the binomial distribution for $H_\mathrm{1}$ gets increasingly skewed and shifted away from the one for $H_\mathrm{null}$.
However, one might select exoplanets for which a biosignature test is performed based on \textit{a priori} available contextual information~\citep{catling2018exoplanet} in order to maximize the science yield of investing additional resources.
For instance, the distribution of $F_\mathrm{NUV}$ in the planet sample can be influenced by the survey strategy, and a targeted sampling approach could favor extreme values. 
We model this by distributing $F_\mathrm{NUV}$ according to different $Beta$ functions and introduce a selectivity parameter $s\in]-1,1[$ such that $F_\mathrm{NUV} \sim Beta(1/10^s,1/10^s)$.
Figure~\ref{fig:semian_selectivity} shows how the probability of obtaining true strong evidence for $H_\mathrm{1}$ scales with selectivity $s$.
For large samples, a high selectivity ($s \sim 1$) can increase the probability of obtaining true strong evidence from $\sim \SI{70}{\percent}$ for $s= 0$ (random uniform distribution) to $> \SI{90}{\percent}$.

\subsection{Survey simulations with \bioverse}\label{sec:results-bioverse}
With \gls{HZ} occupancy as a requirement for abiogenesis, and barring selection biases beyond stellar brightness, the host star distribution of inhabited planets in a simulated transit survey is skewed toward later spectral types.
For a fixed planet sample size, the fraction of inhabited planets is highest for planets orbiting M~dwarfs due to the higher \gls{NUV} fluxes in the \gls{HZ} of these stars (see Figures~\ref{fig:hz_nuv_evo},~\ref{fig:surveys_FGKM}).
Their \gls{NUV} fluxes are generally highest at early times $\lesssim\SI{100}{\mega\year}$.
These host stars, in particular late subtypes, also provide extended periods of increased \gls{NUV} emission that overlap with times when some of these planets occupy the \gls{HZ} (see Figure~\ref{fig:hz_nuv_evo}), our requirement for abiogenesis (see Equation~\ref{eq:hypothesis-uv}).
Thus -- under the UV Threshold Hypothesis -- most inhabited transiting planets in the sample orbit M~dwarfs.

Here, we are interested in the statistical power of a transit survey with a plausible sample selection and size.
In the following, we fix the sample size to \var{N_nautilus} and consider two different survey strategies targeting FGK and M~dwarfs, respectively.
We further investigate the sensitivity of the survey to the a priori unknown threshold \gls{NUV} flux $F_\mathrm{NUV, min}$ and the \rev{probability of life emerging and persisting} $f_\mathrm{life}$.

\subsubsection{Selectivity of simulated transit surveys}
\begin{figure*}
    \script{surveys_FGKM.py}
    \begin{centering}
        \includegraphics[width=\hsize]{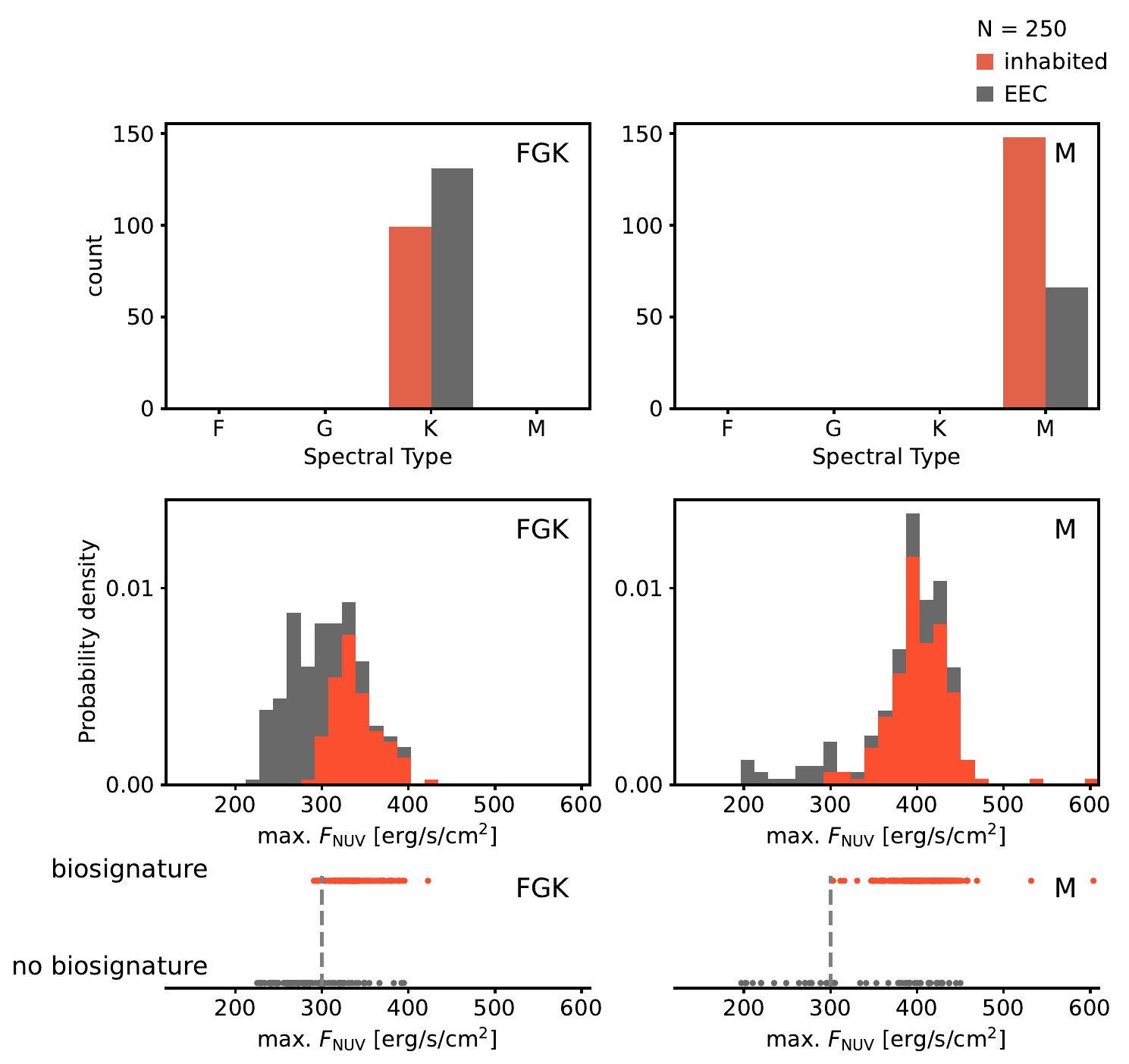}
        \caption{
        Simulated transit surveys targeting FGK and M~stars.\\
        Top: Host stars of all transiting \glspl{EEC} and inhabited planets in a simulated transit survey.
        In the FGK sample, all \glspl{EEC} and all inhabited planets orbit K~dwarfs.
        In an M~dwarf sample of the same size, the fraction of inhabited planets is larger.\\
        Center: Distribution of inferred maximum past \gls{NUV} flux in transit surveys targeting \glspl{EEC} around FGK and M~stars, respectively. The best-fit beta distributions correspond to selectivities of $s_\mathrm{FGK} = $\var{selectivity_FGK} and $s_\mathrm{M} = $\var{selectivity_M}. Red areas show inhabited planets for an abiogenesis rate of $f_\mathrm{life} = \var{f_life}$ and a generic threshold \gls{NUV} flux $F_\mathrm{NUV, min} = \var{NUV_thresh}\,\SI{}{\erg\per\second\per\centi\meter\squared}$.\\
        Bottom: Recovered biosignature detections and non-detections of simulated transit surveys. The dashed line denotes $F_\mathrm{NUV, min}$.}
        \label{fig:surveys_FGKM}
    \end{centering}
\end{figure*}

In Section~\ref{sec:results-semianalytical}, we showed that the probability of obtaining true strong evidence for the hypothesis that life only originates on planets with a minimum past \gls{NUV} flux is sensitive to the distribution of sampled past \gls{NUV} fluxes, i.e., the selectivity of the survey (compare Figure~\ref{fig:semian_selectivity}).
For both surveys targeting M dwarfs and those targeting FGK dwarfs, the maximum \gls{NUV} distribution is rather unimodal.
Applying the approach from Section~\ref{sec:results-semianalytical} of fitting a Beta function to the distribution, we find rather low selectivities (see Figure~\ref{fig:surveys_FGKM}), which is likely detrimental for statistical hypothesis tests.


\subsubsection{Expected biosignature pattern}
A representative recovery of the injected biosignature pattern is shown in Figure~\ref{fig:surveys_FGKM}.
There, we assumed an abiogenesis rate of $f_\mathrm{life} = \var{f_life}$ and a minimum \gls{NUV} flux of $F_\mathrm{NUV, min} = \var{NUV_thresh}\,\SI{}{\erg\per\second\per\centi\meter\squared}$.
All injected biosignatures are assumed to be detected without false positive ambiguity, and the maximum \gls{NUV} flux is estimated from the host star's spectral type and age with an uncertainty corresponding to the intrinsic scatter in the \gls{NUV} fluxes in \citet{Richey-Yowell2023}.
This leads to a distribution of biosignature detections with detections increasingly occurring above a threshold inferred \gls{NUV} flux.
In this example case, the few biosignature detections in the FGK sample lead to a higher evidence 
than in the M dwarf sample
, where the majority of planets are above the threshold \gls{NUV} flux.

\begin{figure}
    \script{inhabited_aafo_nuv_thresh.py}
    \begin{centering}
        \includegraphics[width=\hsize]{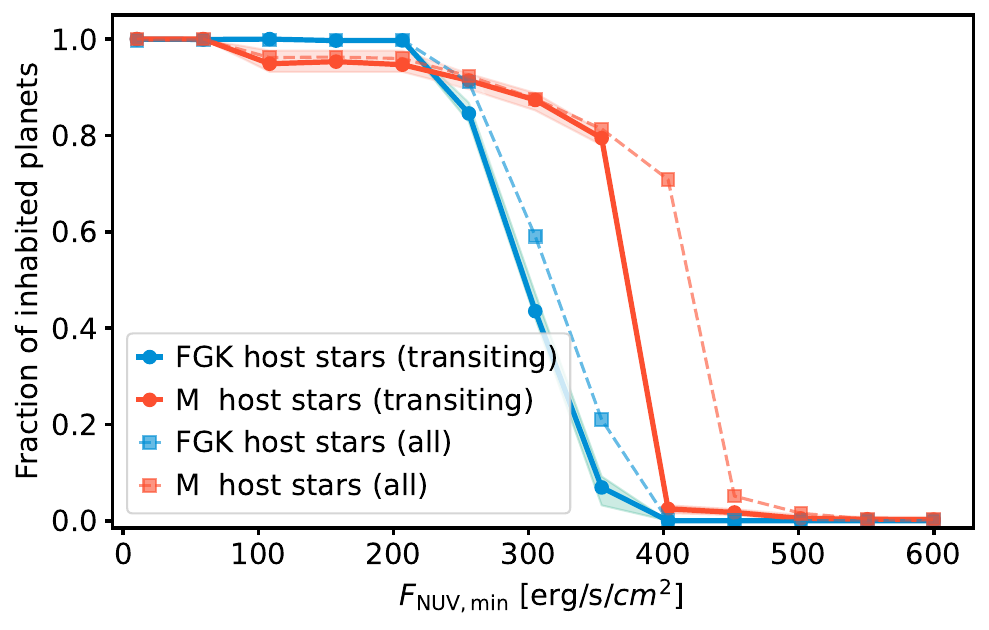}
        \caption{Fraction of inhabited planets for different threshold \gls{NUV} fluxes under the UV~Threshold Hypothesis if the abiogenesis rate $f_\mathrm{life} = 1$.
        For all samples, the fraction of inhabited planets drops sharply with increasing threshold \gls{NUV} flux due to the combined effects of never receiving sufficient \gls{NUV} flux for abiogenesis or receiving it before entering the \gls{HZ}.}
        \label{fig:inhabited_aafo_nuv_thresh}
    \end{centering}
\end{figure}
Figure~\ref{fig:inhabited_aafo_nuv_thresh} shows the fraction of inhabited planets under the UV~Threshold Hypothesis for different threshold \gls{NUV} fluxes and for the limiting case of \rev{a probability for life's emergence and persistence} of $f_\mathrm{life} = 1$.
This fraction decreases sharply with increasing threshold flux, as fewer planets receive sufficient \gls{NUV} flux for abiogenesis.
Another effect responsible for this drop is that some planets receive the required \gls{NUV} flux only before entering the \gls{HZ} -- this is especially likely for M~dwarfs\revv{~\citep{Spinelli2024}}.
For the FGK sample, the fraction of inhabited planets drops at lower threshold fluxes than for the M dwarf sample.

\subsubsection{Statistical power and sensitivity to astrophysical parameters}\label{sec:results-powergrid}
\begin{figure*}
    \script{powergrid.py}
    \begin{centering}
        \includegraphics[width=\hsize]{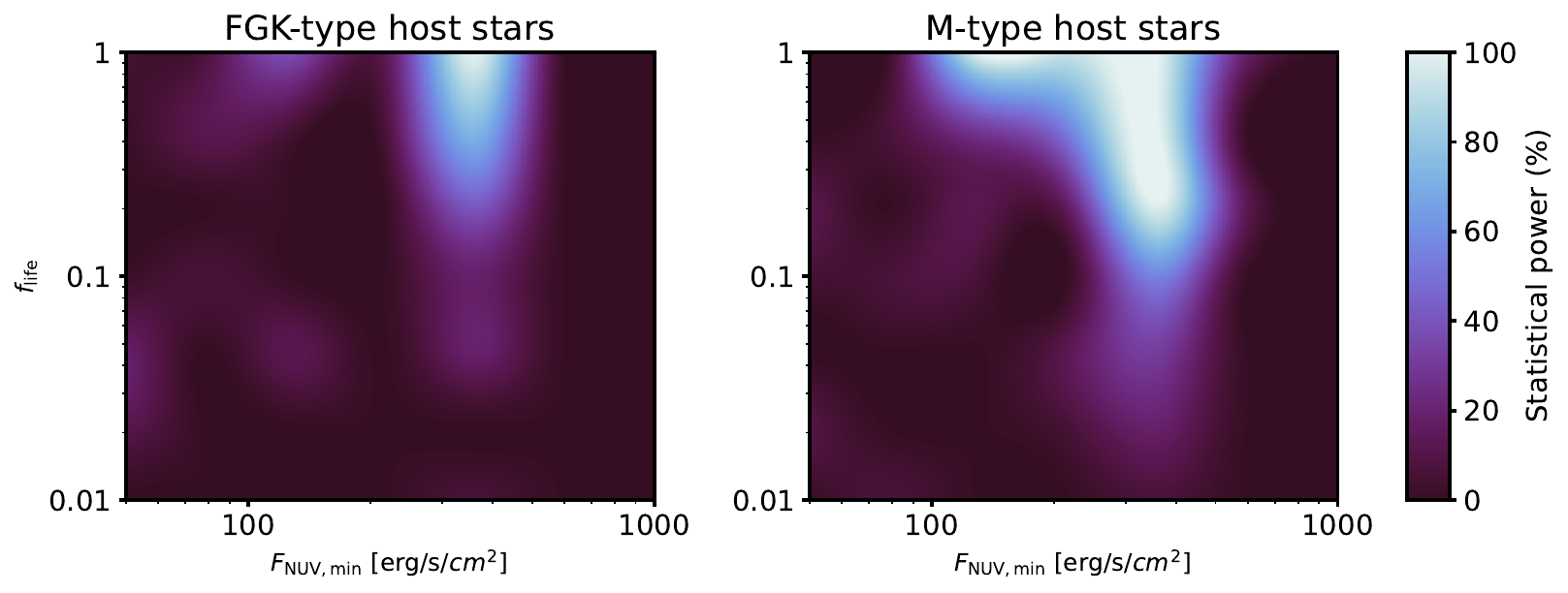}
        \caption{Statistical power as a function of threshold \gls{NUV} flux and abiogenesis rate. Even for a large sample (here: $N=\var{N_nautilus}$), a high statistical power of the transit survey requires high \rev{rates of life emerging and persisting} $f_\mathrm{life}$.
         Intermediate values of $F_\mathrm{NUV, min}$ are more likely to yield strong evidence than extreme values. \rev{For $f_\mathrm{life} \gtrsim \SI{80}{\percent}$}, the sensitivity of the M~dwarf sample extends into the low \gls{NUV} flux end.
         }
        \label{fig:powergrid}
    \end{centering}
\end{figure*}

We now investigate the sensitivity of the achieved statistical power of our default transit survey to the a priori unconstrained threshold \gls{NUV} flux $F_\mathrm{NUV, min}$ and the abiogenesis \rev{and persistence} rate $f_\mathrm{life}$.
Figure~\ref{fig:powergrid} shows the statistical power as a function of these parameters for a sample size of $N=\var{N_nautilus}$.
Values of $F_\mathrm{NUV, min}$ that lie between the extrema of the inferred maximum \gls{NUV} flux increase the achieved statistical power of the survey, as in this case the dataset under the alternative hypothesis $H_1$ differs more from the null hypothesis.
Furthermore, \rev{higher $f_\mathrm{life}$ increase} the evidence for $H_1$.

Parameter space regions with statistical power above \SI{90}{\percent} lie at \rev{$f_\mathrm{life} > 0.5$} and mostly at threshold \gls{NUV} fluxes of $\sim \SIrange{200}{400}{\erg\per\second\per\centi\meter\squared}$.
Notably, the sensitivity of the M~dwarf sample extends into the low \gls{NUV} flux end due to the broader distribution of maximum past \gls{NUV} fluxes in this sample.
Here, the FGK sample is barely sensitive.

\section{Discussion}
\label{sec:discussion}
A key question in the quest to understand the origins of life is which natural processes best explain how living matter spontaneously appears from nonliving matter~\citep[e.g.,][]{Malaterre2022}.
Using astronomical methods, this question will likely not be testable for individual planets but rather for ensembles of planets.
The cyanosulfidic scenario for the origins of life~\citep{Patel2015}, in particular its predicted existence of a minimum \gls{NUV} flux required for prebiotic chemistry, offers an opportunity to test an origins of life hypothesis with a statistical transit survey sampling planets with varying \gls{NUV} flux histories.
In the following, we discuss the prospects of testing the UV~Threshold Hypothesis in light of our results.

\subsection{Sampling strategy for testing a \gls{NUV} flux threshold} 
In Section~\ref{sec:results-semianalytical}, we show that testing the UV~Threshold Hypothesis suffers from `nuisance' parameters that hamper inference through astronomical observations.
Here, these parameters are the unspecified value of the \gls{NUV} threshold hypothesized to exist under $H_1$, and the unknown probability of detectable life emerging on a habitable planet $f_{\mathrm{life}}$.
While the inference of a planet's entire \gls{UV} flux evolution is difficult~\citep[e.g.,][]{Richey-Yowell2023}, the estimated maximum \gls{NUV} flux that a planet was exposed to may be used as a proxy, at least if one is interested in a minimum threshold flux and makes the assumption that planetary surfaces offer protection against \textit{too high} UV flux.
Indeed, the distribution of the number of planets with detected biosignature in a particular sample of planets with inferred maximum \gls{NUV} values $F_{\mathrm{NUV}}$ depends on both the values of $F_{\mathrm{NUV},min}$ and $f_{\mathrm{life}}$ as shown in equation~\ref{eq:semian:likelihoodH1}.

In our semi-analytical analysis (Section~\ref{sec:results-semianalytical}), we project a possible test performed by a future observer equipped with a sample of exoplanets with derived past maximum \gls{NUV} exposure for which biosignature detection has been attempted.
This is necessarily reductive as this observer will have more knowledge about experimental conditions and will therefore be able to use this information to guide hypothesis testing.
For instance, we have made the choice to consider the total number of detected biosignatures as our summary statistic (Equation~\ref{eq:semian:likelihoodH1}), which is not sufficient to infer $F_{\mathrm{NUV},min}$ and $f_{life}$ separately.
However, by conditioning the Bayes factor to these variables (Equation~\ref{eq:bayes_factor}), we calculate the probability distribution of the Bayesian evidence in favor of $H_1$.
In doing so, we may evaluate how evidence depends on the uncertainty over these unknown parameters in general terms, without assuming which particular test a future observer might actually choose to perform over real data when available.
From this, we can see that target selection can strongly affect the conclusiveness of a future test of the UV~Threshold Hypothesis.

The particular finding that prioritizing extreme values of past \gls{NUV} flux can enhance statistical power likely clashes with observational constraints, as the composition of the subset of planets that we can observe and for which detection of biosignature can be attempted is not independent from their \gls{NUV} flux history.
Hence, for our future observer, selectivity and sample size may be in conflict.
This trade-off can be quantified in terms of expected evidence yield, which we have done in Section~\ref{sec:results-semianalytical}.
Our analysis shows that regardless of selectivity, sample sizes smaller than 50 likely result in inconclusive tests, and that increasing selectivity towards extreme $F_{\mathrm{NUV}}$ offers limited inference gains compared to the uniform case ($s=0$; Figure~\ref{fig:semian_selectivity}).
For larger samples, however, a narrow distribution of $F_{\mathrm{NUV}}$ may prevent inference entirely.
We thus argue that selecting a sample with $F_{\mathrm{NUV}}$ distributed uniformly or emphasizing extreme values should -- barring any practical counterarguments -- be considered in any future attempt at testing the UV~Threshold Hypothesis.
Since the practical implementation of an exoplanet survey can stand in the way of such a selection, the following discussion focuses on the results of our transit survey simulations with \bioverse.


\subsection{How planetary context may constrain the UV~Threshold Hypothesis} 
It comes to no surprise that the success rate for testing the UV~Threshold Hypothesis is sensitive to the sample size of the survey and to the occurrence of life on temperate exoplanets.
As we have shown, the statistical power of this test also depends on the distribution of past \gls{NUV} fluxes in the sample and on the threshold flux.
Optimizing the survey to sample a wide range of \gls{NUV} flux values, particularly at the extremes, can enhance the likelihood of obtaining strong evidence for or against the hypothesis.
Intermediate values of the threshold \gls{NUV} flux are more likely to yield strong evidence than extreme values, as the dataset under the alternative hypothesis $H_1$ differs more from the null hypothesis in this case while still being sufficiently populated.
The threshold flux is, of course, a priori unknown and we cannot influence it.
If, however, better theoretical predictions for the required \gls{NUV} flux for abiogenesis become available~\citep{Rimmer2021}, the survey strategy can be further optimized, for instance by targeting planets that are estimated to have received a \gls{NUV} flux slightly below and above this threshold or by applying a bisection algorithm in a sequential survey~\citep{Fields2023}.

\subsection{An M~dwarf opportunity}
An interesting aspect lies in the distribution of host star properties, as different spectral types probe different past \gls{NUV} flux regimes.
FGK stars \rev{exhibit a relatively} narrow \rev{range} of maximum past \gls{NUV} fluxes in the \gls{HZ}, which may -- depending on the (unknown) threshold \rev{flux} -- limit the diagnostic power of a survey.
\rev{A} pure FGK sample \rev{would} only be sensitive to \rev{flux} thresholds \rev{in the range of} $\sim\SIrange{200}{400}{\erg\per\second\per\centi\meter\squared}$.
\rev{A detection of biosignatures in such a sample would likely only suggest that either low \gls{NUV} fluxes are sufficient for abiogenesis or that an alternative abiogenic pathway may be at play~\citep[e.g.,][]{Westall2018}.
Conversely, a lack of biosignatures in a sufficiently large FGK sample could indicate that the actual threshold is higher than the maximum past flux levels reached by FGK hosts.
We further note that under the UV~Threshold Hypothesis, the fraction of inhabited planets in an FGK sample declines rapidly for increasing threshold fluxes, as shown in Figure~\ref{fig:inhabited_aafo_nuv_thresh}.
In addition, if the timescale for the emergence of life is $\gg \SI{1}{\mega\year}$, the fraction of inhabited planets in an FGK sample may be negligible (see Section~\ref{sec:long-origins}).
}

On the other hand, M~dwarfs show a wider distribution of maximum past \gls{NUV} fluxes in their \glspl{HZ}.
While old M~dwarfs can be considered low-UV environments, a significant fraction of them emit high \gls{NUV} fluxes into their \gls{HZ} during their early stages, in particular later subtypes\revv{~\citep{Richey-Yowell2023,Spinelli2024}}.
This will help to test the high \gls{NUV} flux end of the UV Threshold Hypothesis; a higher occurrence of biosignatures here would support the hypothesis that a higher \gls{NUV} flux is favorable or necessary for life.
At the same time, a fraction of host stars in our M~dwarf sample extends it to lower maximum past \gls{NUV} fluxes, enabling tests of the low \gls{NUV} flux end of the hypothesis.
The higher and more variable \gls{NUV} fluxes in M~dwarfs thus increase the likelihood of obtaining strong evidence for or against the UV~Threshold Hypothesis.

\revv{It was found that the ``abiogenesis zone'', defined as the region around a star where quiescent UV flux is sufficient to drive the prebiotic photochemistry necessary for building RNA precursors, does not necessarily overlap with the liquid water \gls{HZ} for stars cooler than K5 ($T_{\mathrm{eff}} \lesssim \SI{4400}{\kelvin}$), implying that such stars may be less favorable for initiating life via UV-driven chemistry~\citep{Rimmer2018}.
This conclusion, however, is based on a steady-state view of stellar UV fluxes and habitable zone boundaries.
In our work, we addressed this limitation by explicitly incorporating stellar evolution, including the high-UV early phase of M~dwarfs.
Our results suggest that planets orbiting M dwarfs may still be viable sites for abiogenesis under UV-dependent pathways.
In fact, the} combination of a lack of \gls{UV} radiation today, which makes biosignature gases more detectable~\citep{Segura2005}, and a UV-rich past that may have enabled abiogenesis could make M~dwarfs the preferred targets for biosignature searches.
We note that relevant mission concepts, such as the Large Interferometer for Exoplanets~\citep[\life,][]{Quanz2022,Glauser2024}, include M-dwarf systems among their primary targets~\citep{Kammerer2018,Carrion-Gonzalez2023}.
Our findings underscore the importance of constraining the \gls{UV} emission profiles of \gls{EEC} host stars throughout their evolutionary stages to assess the viability of M-dwarf planets as testbeds for theories on the origins of life~\citep{Rimmer2021,Ranjan2023a}.

\subsection{Sensitivity to astrophysical parameters} 
Our \bioverse\ simulations that take into account exoplanet demographics, the evolution of habitability and \gls{NUV} fluxes, and observational biases show that not only the likelihood of a conclusive test of the UV~Threshold Hypothesis, but also the likelihood of successful biosignature detection itself is extremely sensitive to the threshold \gls{NUV} flux if the hypothesis is true.
Even if all biosignatures can be detected and the nominal \rev{rate of life's emergence and persistence} is very high, say $f_\mathrm{life} = 1$, under the condition that prebiotic chemistry requires a minimum \gls{NUV} flux \textit{and} liquid water, if the threshold flux turns out to be high the probability of finding life on a randomly selected planet may be very low.
As we showed, for high required fluxes the two requirements of simultaneous \gls{HZ} occupancy and sufficient \gls{NUV} flux conspire to diminish the fraction of inhabited planets in the sample.
Taking the inferred fluxes from \citet{Richey-Yowell2023} at face value (but taking into account intrinsic scatter), a minimum required \gls{NUV} flux of $\gtrsim\SI{400}{\erg\per\second\per\centi\meter\squared}$ reduces the fraction of inhabited planets to below $\sim\SI{1}{\percent}$.
This not only calls for a large sample size and a targeted sample selection preferring high expected past \gls{NUV} fluxes, but also highlight the necessity of continued theoretical and experimental research into the role of \gls{UV} radiation in prebiotic chemistry~\citep{Ranjan2017b,Rimmer2018,Rimmer2021}.


\subsection{Contextual support for potential biosignature detections} 
The predicted interplay of \gls{NUV} flux and \gls{HZ} occupancy in enabling abiogenesis via the cyanosulfidic scenario could in principle be used to add or remove credibility from a tentative biosignature detection.
For example, with a strong \rev{prior} belief that this scenario is the only viable one for the origins of life \rev{(see Section~\ref{sec:discussion-bayes})}, a biosignature detection on a planet orbiting a strongly UV-radiating star may add credibility to the detection.
Conversely, a biosignature detection on a planet estimated to have received very little UV~radiation would increase the likelihood of a false positive detection.
On the other hand, should the detection in the latter case be confirmed, it could be used to falsify the UV~Threshold Hypothesis.

Our simulations find no clear criterion for the credibility of a biosignature detection based on spectral type of the host star, as both FGK and M~dwarf samples show similar maximum past \gls{NUV} flux distributions.
\rev{While there is a significant preference for M~dwarfs in the case of a long timescale for abiogenesis (see Appendix \ref{sec:long-origins}
), the fraction of inhabited planets in the M~dwarf sample drops at similar threshold fluxes as in the FGK sample if the timescale for abiogenesis is short }(see Section~\ref{sec:results-bioverse}).
A potentially inhabited planet's host star spectral type may thus not be a strong indicator for the credibility of a biosignature detection in the context of the UV~Threshold Hypothesis.

\subsection{Overall prospects for testing the UV~Threshold Hypothesis} 
Our results show that the UV~Threshold Hypothesis is testable with potential future exoplanet surveys, but that the success of such a test depends on the sample size, the distribution of past \gls{NUV} fluxes, and several unknown astrophysical nuisance parameters.
Even under idealized conditions, obtaining strong evidence for or against the hypothesis likely requires sample sizes on the order of 100 (see Section~\ref{sec:results-semianalytical}).
This is true for a future transit survey, the specifics of which we have reflected in our \bioverse\ simulations (see Section~\ref{sec:results-bioverse}).
However, we have shown that the impacts from the combined requirements of the UV~Threshold Hypothesis on the fraction of inhabited planets in a sample are comparable in the non-transiting case.

Given the challenging nature of detecting and characterizing small (Earth-sized) exoplanets, most exoplanet mission concepts currently under development or considered lack the potential for characterizing large enough samples.
Ground-based 25--40-meter class extremely large telescopes are expected to have the capabilities to detect biosignatures on exoplanets like Proxima Centauri~b~\citep[e.g.,][]{Wang2017,Hawker2019,Zhang2024,Vaughan2024}.
\citet{Hardegree-Ullman2025} used \bioverse\ to determine potential yields for a 10-year direct imaging and high-resolution spectroscopy survey of O$_2$ on the Giant Magellan Telescope (GMT) and on the Extremely Large Telescope (ELT) and found that between 7 and 19 habitable zone Earth-sized planets could be probed for Earth-like oxygen levels.
Such a sample is too small to test the UV Threshold Hypothesis, but it may be synergistic with other detection methods.

The \gls{HWO} is expected to characterize a sample of $\sim 25$ Earth analogs~\citep{Mamajek2023,Tuchow2024}.
Depending on the technical design, \life\ is expected to target 25--80 \glspl{EEC}~\citep{Kammerer2018,Quanz2022}, which could be just sufficient to constrain the UV~Threshold Hypothesis.

\rev{If the rate of life's emergence and persistence $f_\mathrm{life}$ is at a \SI{1}{\percent} level or lower, no currently envisioned future exoplanet mission has projected sample sizes sufficient to test the UV~Threshold Hypothesis.
One possible} exception is the Nautilus Space Observatory concept~\citep{Apai2019,Apai2022}.
Nautilus aims to characterize up to $\sim 1000$ \gls{EEC} via transmission spectroscopy, building on an innovative optical technology.
To guide the definition of future biosignature surveys, it is important to refine predictions on the role of \gls{UV} radiation in prebiotic chemistry with both theoretical and experimental work.

\subsection{Caveats}
Our work is based on a number of assumptions and simplifications that may affect the results and conclusions.
We discuss some of these caveats here.

\rev{\subsubsection{The UV~Threshold Hypothesis as a narrow step function}
}
\rev{A key aspect of the UV Threshold Hypothesis is the proposed step-function dependence of abiogenesis likelihood on UV flux.
This approach stems from the constraints governing photochemical pathways, which exhibit a threshold behavior: below a certain flux, competing thermal reactions dominate, preventing abiogenesis, while above the threshold, UV photochemistry proceeds at sufficient rates, and other stochastic processes become rate-limiting.}

\rev{\citet{Ranjan2017b} speculated that UV photochemistry might be rate-limiting for abiogenesis, particularly on planets orbiting M dwarfs, due to their low baseline UV fluxes.
This could delay abiogenesis by orders of magnitude, resulting in a continuous dependence of abiogenesis likelihood on UV flux.
However, recent studies challenge this view for the cyanosulfidic scenario.
\citet{Rimmer2021} calculated photochemical timescales on early Earth at 180--300 hours (7.5--12.5 days), significantly shorter than the timescale for stochastic geological events also required for the scenario~\citep{Rimmer2023}.
Even with 1000x slower photochemistry on M-dwarf planets, prebiotic photochemistry like sulfite photolysis would still occur on a geologically negligible timescale of 20--30 years, meaning that the photochemistry is unlikely to be rate limiting compared to stochastic geological processes, justifying the step-function model.
}

\revv{We note that the photochemical timescale does not include subsequent stages such as assembly into functional biomolecules, protocell formation, or the emergence of self-replicating systems -- each of which may require significantly more time.
However, in the context of the cyanosulfidic scenario, UV irradiation is needed only for this initial synthetic step, and later stages can proceed under different environmental conditions, potentially long after UV levels have declined.
Moreover, abiogenesis may require many ``failed trials''.
We effectively control for this uncertainty by requiring that both sufficient UV flux and habitable conditions persist for a minimum duration $\Delta T_\mathrm{min}$.
We further require that the planet remains in the habitable zone at the time of observation, thereby capturing both the origination and persistence phases encapsulated by $f_\mathrm{life}$.
}

\rev{Nonetheless, alternative abiogenesis pathways or combinations of pathways may exhibit continuous or mixed dependencies on UV flux.}
\revv{Future work} should explore UV dependencies across other scenarios to refine predictions for biosignature distributions and testable hypotheses.

\subsubsection{Existence of an atmosphere-crust interface}
By its nature, cyanosulfidic scenario relies on rock surfaces exposed to the planetary atmosphere.
Water worlds that have their entire planetary surface covered by oceans contradict this requirement and do not allow for the wet-dry cycling inherent to this origin of life scenario.
The competition of tectonic stress with gravitational crustal spreading~\citep{Melosh2011} sets the maximum possible height of mountains, which in the solar system does not exceed $\SI{\sim 20}{\kilo\meter}$.
Such mountains will be permanently underwater on water worlds.
Another impediment to wet-dry cycles may be tidal locking of the planet as it stalls stellar tide-induced water movement and diurnal irradiation variability~\citep[e.g.,][]{Ranjan2017b}.
However, recent dynamical models suggest tidally locked planets to undergo rapid drift of their sub-stellar point~\citep{Revol2024}.

\subsubsection{Stellar flares}
Our assumptions on past \gls{UV} flux neglect the contribution of stellar flares, which may be hypothesized as an alternative source of \gls{UV} light~\citep{Buccino2007,Ranjan2017c}.
This concerns mainly ultracool dwarfs, due to their low quiescent emission and high pre-main sequence stellar activity~\citep{Buccino2007,West2008}.
However, recent work indicates that the majority of stars show inadequate activity levels for a sufficient contribution through flares~\citep{Glazier2020,Ducrot2020,Guenther2020}.

\subsubsection{Atmosphere transmission}
We do not take into account absorption of \gls{UV} radiation by the planetary atmosphere.
Theoretical work suggests that the atmosphere of prebiotic Earth was largely transparent at \gls{NUV} wavelengths with the only known source of attenuation being Rayleigh scattering~\citep{Ranjan2017,Ranjan2017c}.
We thus approximated surface \gls{UV} flux using top-of-atmosphere fluxes.
If there are planets in a sample that do not have a transparent atmosphere at \gls{NUV} wavelengths and require higher fluxes for abiogenesis, the fraction of inhabited planets in the sample will be lower.
However, these planets will not pollute the below-threshold subsample, as they will not be able to host life under the UV~Threshold Hypothesis.
Exoplanet surveys focusing on highly irradiated planets offer an opportunity to constrain the typical oxidation state of rocky exoplanets, providing insights into the average composition of their secondary atmospheres~\citep{Lichtenberg2024}.
This is particularly relevant for prebiotic worlds, as varying oxidation states significantly perturb the classical habitable zone concept~\citep{Nicholls2024} and also influence surface \gls{UV} levels through changing atmospheric transmission.
Optimally, the atmospheric composition of young rocky protoplanets will be probed to constrain the possible range of atmospheric and mantle oxidation states during early planetary evolution by future direct imaging concepts~\citep{Cesario2024}.


\rev{
\subsubsection{Other mechanisms regulating habitability and abiogenesis}
Our study focuses primarily on two factors that may regulate the emergence and persistence of life on a planet: the \gls{NUV} flux and the viability of liquid water, which provide a testable framework for assessing biosignature distributions.
Of course, planetary habitability and abiogenesis depend on a broader range of physical and chemical conditions.
For one, a planet must retain an atmosphere capable of supporting liquid water and surface chemistry.
Atmospheric escape processes have been studied in detail, and their occurrence is supported by planet formation models~\citep[e.g.,][]{Owen2013,Schlichting2014,Ginzburg2016,Mordasini2020,Burn2024} as well as exoplanet demographics~\citep[e.g.,][]{Owen2019,Bergsten2022,Rogers2021}.
Population synthesis studies demonstrate that models of atmospheric escape can explain key statistical features of the observed exoplanet population~\citep{Rogers2020,Emsenhuber2021b,Schlecker2021b,Burn2024}.}

\rev{
Planets around M~dwarfs, in particular, may be subject to significant atmospheric loss due to the extended high-luminosity pre-main-sequence phase of their host stars}\revv{, and stellar winds can play a major role in stripping secondary atmospheres ~\citep[e.g.,][]{Luger2015,Dong2018,Garcia-Sage2017}.
Recent works~\citep[e.g.,][]{Coy2024,Luque2024} interpret JWST eclipse measurements as growing evidence for the absence of substantial atmospheres on M-dwarf rocky exoplanets.
If these planets indeed lack atmospheres, their habitability would be primarily dictated by the rate of atmospheric escape~\citep[e.g.,][]{Owen2020}.
However, alternative interpretations remain viable: \citet{Ducrot2024} and \citet{Hammond2025} recently demonstrated that current JWST data cannot definitively distinguish between a bare-rock scenario and an atmosphere composed primarily of N$_2$-CO$_2$-H$_2$O.
This underscores the need for future observations and modeling efforts to constrain atmospheric retention and escape processes more robustly.}

\rev{Beyond atmospheric escape, internal heating from tidal forces or radioactive decay can extend or constrain the limits of planetary surface habitability~\citep[e.g.,][]{Barnes2013,Oosterloo2021}.
Tidal effects are especially relevant for \gls{HZ} M~dwarf planets, where strong stellar interactions can drive internal heating, loss of water or an atmosphere, or runaway greenhouse conditions.
Tidal locking may also create extreme climate zones, challenging habitability.}
\rev{While not modeled here, the factors outlined above may offer directions for future work.}

\subsubsection{\rev{Bayesian Evidence and the Influence of Priors}}\label{sec:discussion-bayes}
\rev{Our semi-analytical analysis (Section~\ref{sec:results-semianalytical}) employs Bayes factors to quantify the evidence in favor of or against the UV~Threshold Hypothesis that future observations may provide.}

\rev{
Using the Bayes theorem
to estimate posterior probabilities would allow for a more complete assessment by quantifying information gain and integrating prior knowledge into hypothesis testing.
In practice, implementation of a full Bayesian inference is often hindered by the subjectivity of prior distributions.
Here, we thus chose to provide merely a general assessment of the potential evidence yield from future observations.
In essence, we addressed the question: \textit{By how much does a particular observation tip the scale between the UV~Threshold Hypothesis and the null hypothesis?}
}

\section{Conclusions}
\label{sec:conclusions}
We propose that specific origins-of-life scenarios may leave a detectable imprint on the distribution of biosignatures in exoplanet populations.
We have investigated the potential of upcoming exoplanet surveys to test the hypothesis -- motivated by the cyanosulfidic origins-of-life scenario -- that a minimum past \gls{NUV} flux is required for abiogenesis.
To this end, we first employed a semi-analytical Bayesian analysis to estimate probabilities of obtaining strong evidence for or against this hypothesis.
We then used the \bioverse\ framework to assess the diagnostic power of realistic transit surveys, taking into account exoplanet demographics, time-dependency of habitability and \gls{NUV} fluxes, observational biases, and target selection.

Our main findings are:
\begin{enumerate}
    \item The UV Threshold Hypothesis of the cyanosulfidic scenario for the origins of life should lead to a correlation between past \gls{NUV} flux and current occurrence of biosignatures that may be observationally testable.
    \item The required sample size for detecting this correlation depends on the abiogenesis rate on temperate exoplanets and the distribution of host star properties in the sample; in particular their maximum past \gls{NUV} fluxes.
    Samples smaller than 50 planets are unlikely to yield conclusive results.
    \item Under the UV~Threshold Hypothesis, the fraction of inhabited planets in a transit survey is sensitive to the threshold \gls{NUV} flux and is expected to drop sharply for required fluxes above a few hundred~\SI{}{\erg\per\second\per\centi\meter\squared}.
    \item If the predicted \gls{UV}~correlation exists, obtaining strong evidence for the hypothesis is likely ($\gtrsim \SI{80}{\percent}$) for sample sizes $\geq 100$ if the abiogenesis rate is high ($\gtrsim \SI{50}{\percent}$) and if no very high \gls{NUV} fluxes are required.
    A survey strategy that targets extreme values of inferred past \gls{NUV} irradiation increases the diagnostic power.
    \item Samples of planets orbiting M~dwarfs overall yield higher chances of successfully testing the UV~Threshold Hypothesis.
          They may also be more likely to yield biosignature detections under this hypothesis\rev{, in particular if the origins timescale is long}.


\end{enumerate}

Overall, our work demonstrates that future exoplanet surveys have the potential to test the hypothesis that a minimum past \gls{NUV} flux is required for abiogenesis.
More generally, we found that models of the origins of life provide hypotheses that may be testable with these surveys.
Conducting realistic survey simulations with representative samples is important to identify testable science questions, support trade studies, help define science cases for future missions, and guide further theoretical and experimental work on the origins of life.
Our work highlights the importance of understanding the context in which a biosignature detection is made, which can not only help to assess the credibility of the detection but also to test competing hypotheses on the origins of life on Earth and beyond.

\begin{acknowledgments}
\section*{Acknowledgments}
    The authors thank Kevin Heng, Dominik Hintz, Dominika Itrich, Chia-Lung Lin, and Rhys Seeburger for insightful discussions.
    We thank the anonymous referee for providing constructive critical feedback that helped to improve this manuscript.
    This material is based upon work supported by the National Aeronautics and Space Administration under Agreement No. 80NSSC21K0593 for the program ``Alien Earths''.
    The results reported herein benefited from collaborations and/or information exchange within NASA’s Nexus for Exoplanet System Science (NExSS) research coordination network sponsored by NASA’s Science Mission Directorate.
    This work has made use of data from the European Space Agency (ESA) mission \gaia\ (\url{https://www.cosmos.esa.int/gaia}), processed by the \gaia\ Data Processing and Analysis Consortium (DPAC, \url{https://www.cosmos.esa.int/web/gaia/dpac/consortium}). Funding for the DPAC has been provided by national institutions, in particular the institutions participating in the \gaia\ Multilateral Agreement.
    T.L. was supported by the Branco Weiss Foundation, the Netherlands eScience Center (PROTEUS project, NLESC.OEC.2023.017), and the Alfred P. Sloan Foundation (AEThER project, G202114194).
\end{acknowledgments}

\section*{Author contributions}
M.S., D.A., and S.R.\ conceived the project, planned its implementation, and interpreted the results.
M.S.\ developed the planetary evolution component to \bioverse, carried out the hypothesis tests and statistical analyses, and wrote the manuscript.
D.A.\ leads the ``Alien Earths'' program through which this project is funded, helped to guide the strategy of the project, and provided text contributions.
A.A.\ carried out the semi-analytical computations regarding the correlation of past UV flux and biosignature occurrence.
S.R.\ advised on planetary \gls{NUV} flux evolution and the cyanosulfidic scenario of the origins of life.
R.F.\ wrote the initial draft of the Introduction and advised on the evolutionary biology aspects of the project.
K.H.-U.\ contributed to the \bioverse\ software development and simulations.
T.L.\ supported the selection of testable hypotheses and provided text contributions to the initial draft.
S.M.\ advised on the scope of the project and supported the selection of testable hypotheses.
All authors provided comments and suggestions on the manuscript.


\section*{Reproducibility}
All code required to reproduce our results, figures, and this article itself is available at \url{https://github.com/matiscke/originsoflife}\rev{, and the repository has been archived on Zenodo at \url{https://zenodo.org/records/15022723}.}

\appendix

\section{\rev{Photon Number Fluxes and Energy Fluxes}}\label{appendix:photon_vs_energy_flux}
\begin{figure*}[htb]
    \centering
    \includegraphics[width=.9\textwidth]{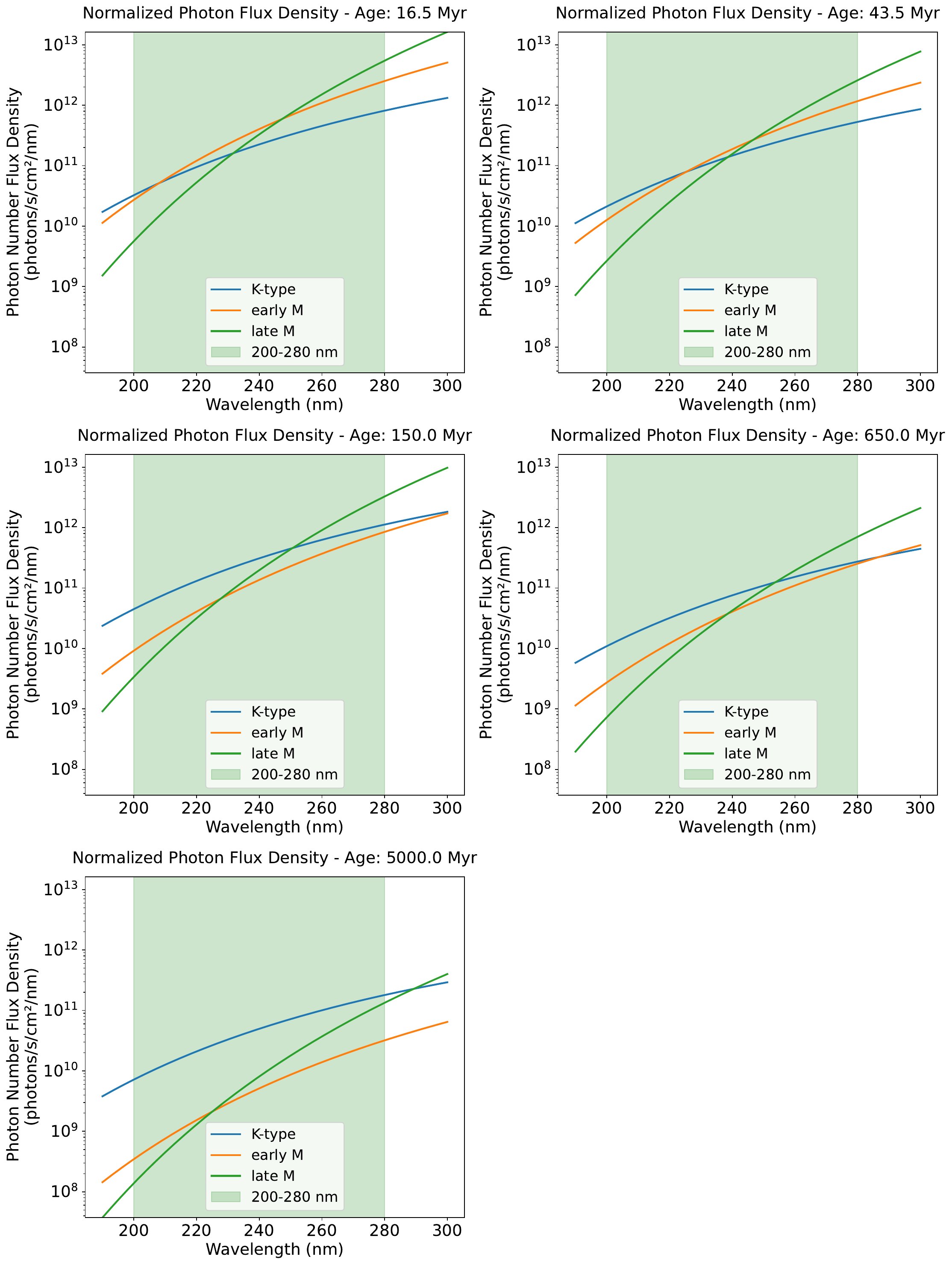}
    \caption{\revv{Photon flux densities of K-type, early M-type, and late M-type stars at different ages.
    All flux densities are normalized to the observed \gls{NUV} fluxes from \citet{Richey-Yowell2023}.
    The green area indicates the \SIrange{200}{280}{\nano\meter} wavelength range used to calculate energy fluxes.
    }}
    \label{fig:normalized_photon_seds}
\end{figure*}

\rev{The \gls{NUV} fluxes we used in our simulations are given in units of energy flux, i.e., \si{\erg\per\second\per\centi\meter\squared}, due to their availability across different spectral types and evolutionary stages of stars.
Here, we assess the robustness of our conclusions when considering photon number fluxes instead.
}
\rev{We first computed blackbody spectral energy distributions (SEDs) for representative effective temperatures corresponding to K-type, early M-type, and late M-type stars, and across the evolutionary stages considered in the main text.
While blackbody SEDs are not necessarily representative of M-dwarf SEDs in the \gls{NUV}~\citep{Seager2013,Rugheimer2015}, the spectrally resolved evolution of M-dwarf SEDs as a function of age is not known; we therefore adopt this simplified prescription for the purpose of this sensitivity test.
We then normalized these SEDs to align with the observed UV fluxes from \citet{Richey-Yowell2023} and used them to calculate mean number flux densities (photons \si{\per\centi\meter\squared\per\second\per\nano\meter}) in the \SIrange{200}{280}{\nano\meter} wavelength range }\revv{(see Figure~\ref{fig:normalized_photon_seds})}.

\begin{figure*}[htb]
    \centering
    \includegraphics[width=\textwidth]{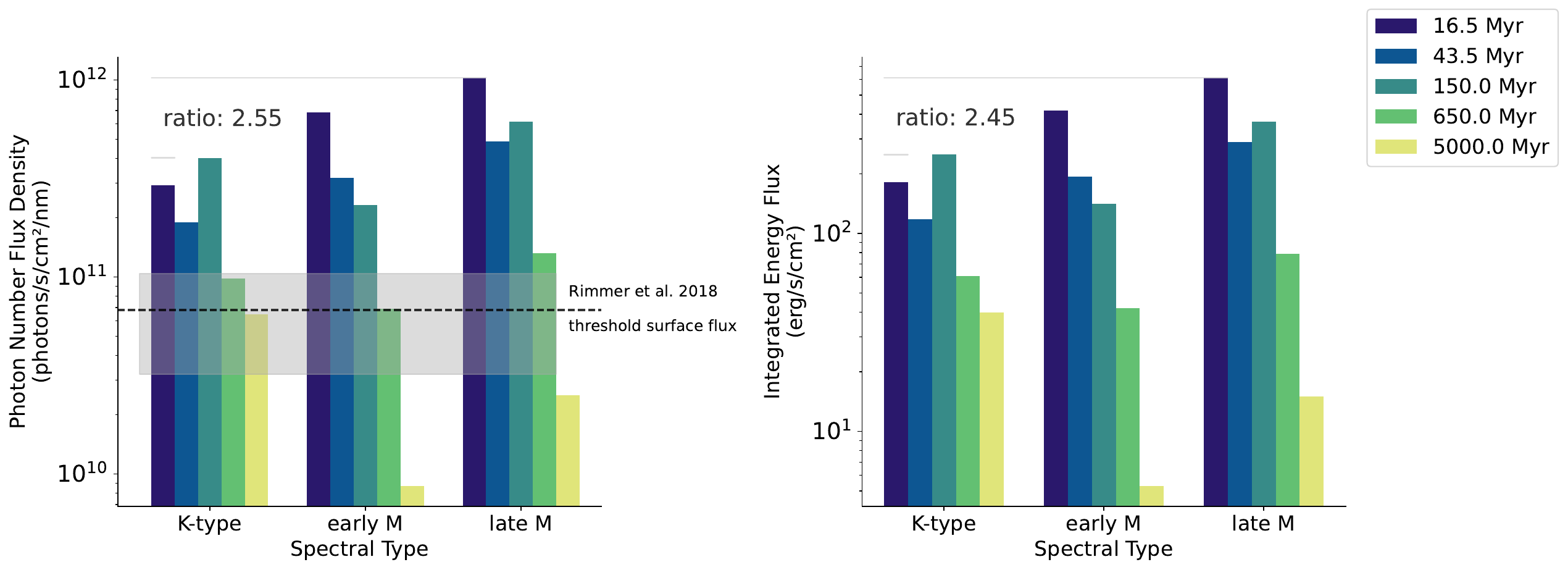}
    \caption{\rev{Comparison of \gls{NUV} flux evolution in terms of photon number flux and energy flux for K-type, early M-type, and late M-type stars.
    In both regimes, later spectral types exhibit higher maximum values.
    The ratio of the maximum past \gls{NUV} fluxes in K dwarfs and late M dwarfs is slightly larger in the photon number flux regime compared to the energy flux regime, suggesting that our conclusions regarding the impact of host star spectral type on the UV Threshold Hypothesis are conservative.
    }}
    \label{fig:combined_fluxes}
\end{figure*}
\rev{We found that the derived flux densities for all considered spectral types exceed the threshold estimated by \citet{Rimmer2018} during earlier stages of stellar evolution\revv{~\citep[Figure~\ref{fig:combined_fluxes}, also see][]{Spinelli2024}}.
Notably -- as in the case of energy fluxes -- later spectral types exhibit higher maximum past \gls{NUV} number flux densities }\revv{due to the steeper rise of the blackbody SEDs at shorter wavelengths}.

\revv{We also calculated } \rev{the ratio of the maximum past \gls{NUV} fluxes in K dwarfs and late M dwarfs in both photon number flux and energy flux regimes.
We found that this ``dynamic range'' is slightly larger in the photon number flux regime (\var{photon_flux_ratio}) compared to the energy flux regime (\var{integrated_flux_ratio}).
This suggests that our use of energy flux may slightly underestimate the variations in \gls{NUV} exposure across different stellar types, rendering our conclusions regarding the impact of host star spectral type on the UV Threshold Hypothesis conservative.
Our results are thus robust when transitioning between energy flux and photon number flux representations.}

\section{\rev{Alternative priors for the abiogenesis rate}}
\label{sec:appendix-flife-prior}
\begin{figure*}
    \begin{centering}
        \includegraphics[width=\hsize]{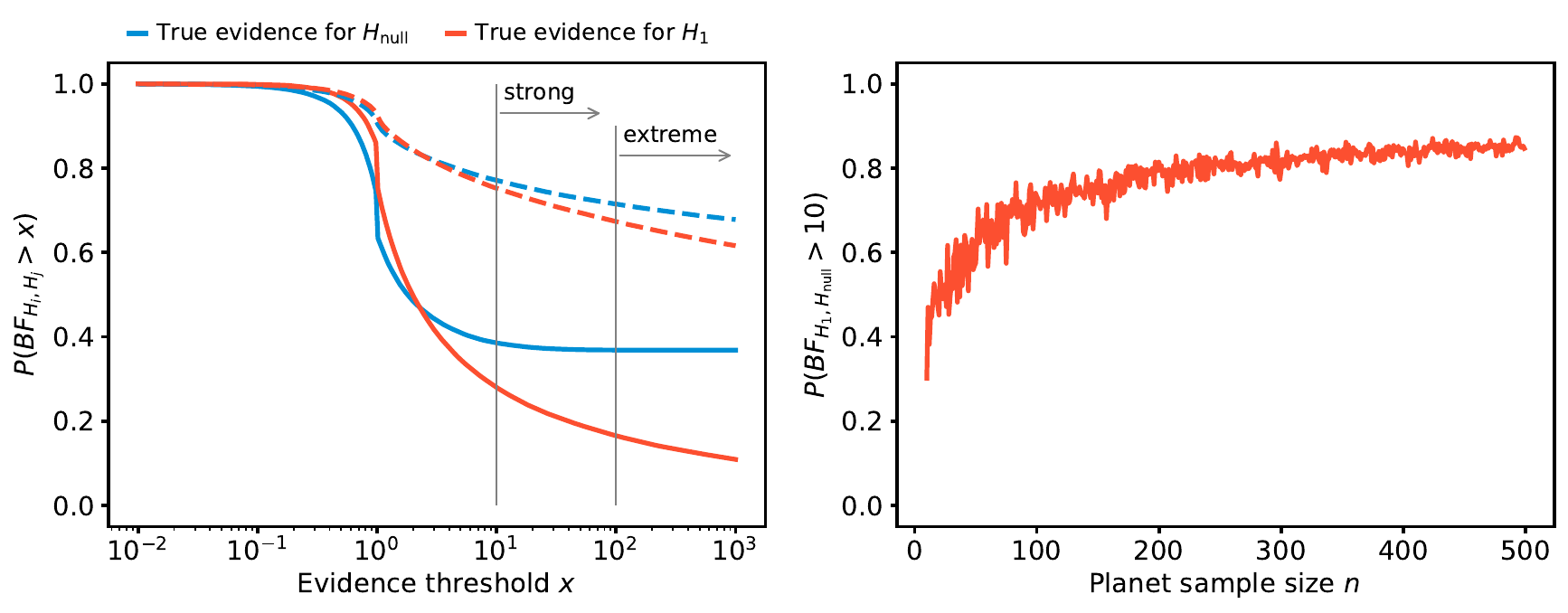}
        \caption{\rev{As Figure~\ref{fig:semian_true_evidence}, but using a uniform prior for the abiogenesis rate $f_\mathrm{life}$.
        The probabilities of obtaining strong evidence for or against the UV~Threshold Hypothesis are overall higher than for the log-uniform prior.
        }
        }
        \label{fig:semian_true_evidence_uniform}
    \end{centering}
\end{figure*}

\rev{In the main text, we used a log-uniform prior \revv{over the range $[10^{-15}, 1]$} for the probability of life emerging and persisting $f_\mathrm{life}$ in order to reflect our ignorance about this parameter and its order of magnitude.
We consider this a reasonable choice, as it is agnostic about the scale of $f_\mathrm{life}$ and assigns equal prior probability to all orders of magnitude.
Nevertheless, we repeated our semi-analytical analysis for the more optimistic choice of a prior that is \textit{uniform} in $f_\mathrm{life}$.
Figure~\ref{fig:semian_true_evidence_uniform} shows the resulting probabilities of obtaining strong evidence for or against the UV~Threshold Hypothesis under this prior.
While the resulting trends are qualitatively similar, the uniform prior yields overall higher probabilities of such a conclusive test for all sample sizes.
}

\section{\rev{Impact of a Longer Origins Timescale}}\label{sec:long-origins}
\rev{
While evidence from Earth's history suggests that life may have emerged early~\citep{Mojzsis2001, Dodd2017}, there is no strong empirical basis to assume that this timescale is representative of habitable planets in the Universe.
The emergence of life could be a slow and rare process, possibly requiring much longer periods \citep{Lineweaver2002, Spiegel2012}.
}

\begin{figure}
    \centering
    \includegraphics[width=0.5\textwidth]{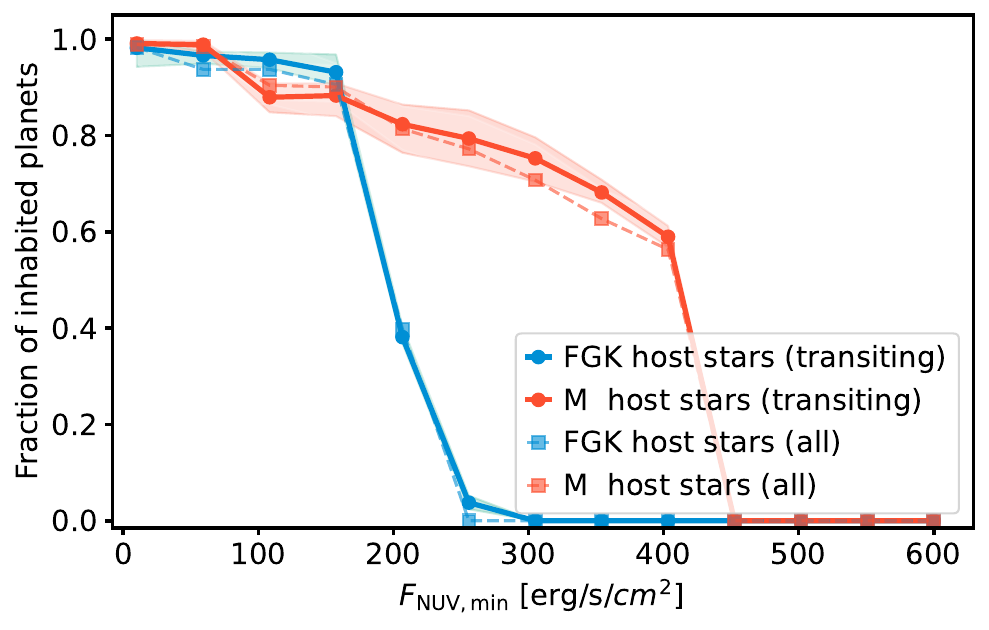} 
    \caption{\rev{Fraction of inhabited planets as a function of NUV threshold flux for $\Delta T_{\mathrm{min}} = 100$ Myr. The decrease in inhabited planets is more pronounced for FGK stars, narrowing the viable range of UV Thresholds compared to shorter origins timescales.}}
    \label{fig:inhabited_dT100}
\end{figure}

\begin{figure*}
    \centering
    \includegraphics[width=0.9\linewidth]{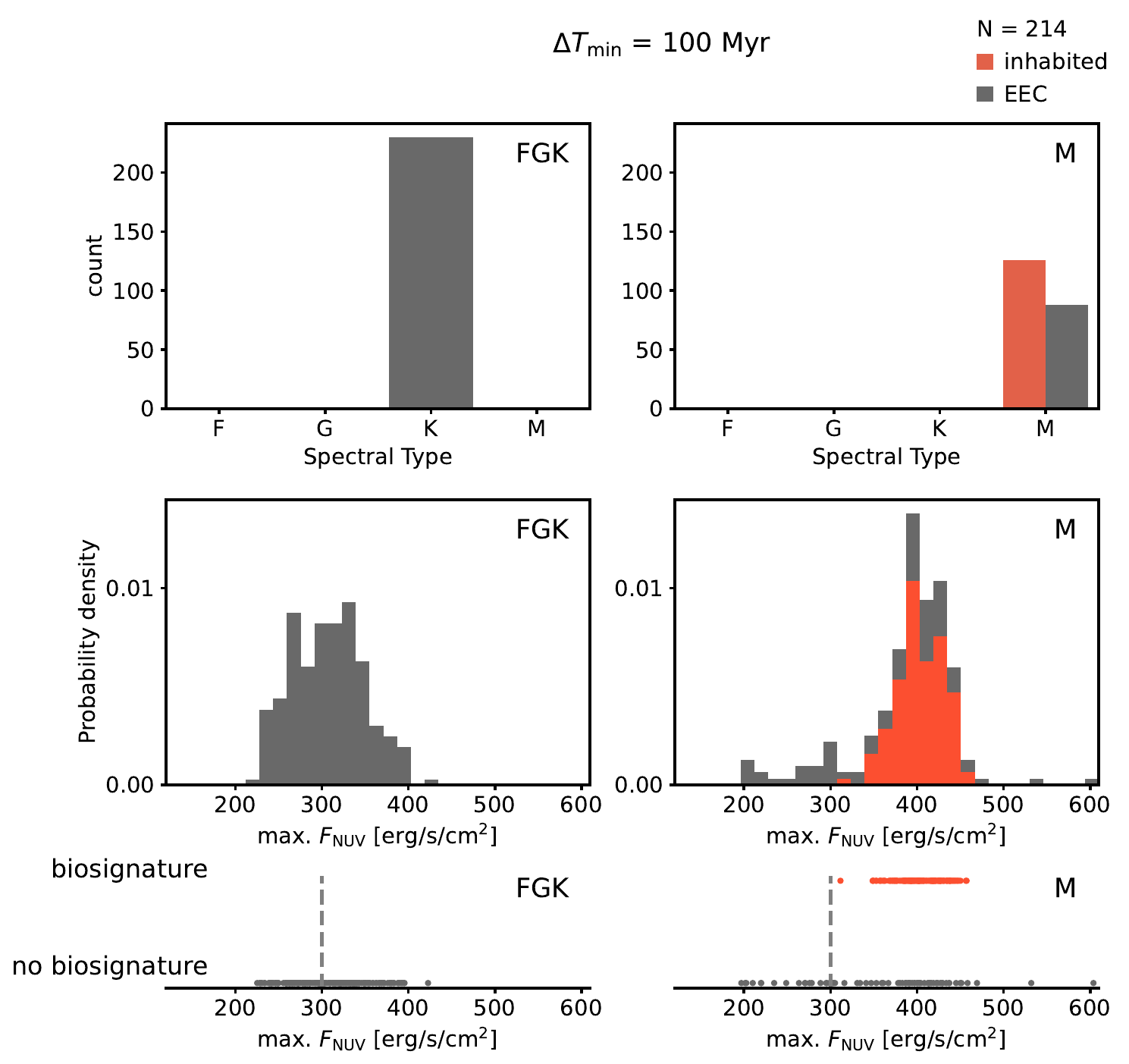}
    \caption{\rev{Simulated survey results under the assumption of $\Delta T_{\mathrm{min}} = 100$ Myr. The fraction of inhabited FGK planets is reduced to zero, while the fraction of inhabited M dwarf planets remains largely unaffected.}}
    \label{fig:surveys_dT100}
\end{figure*}

\begin{figure*}
    \centering
    \includegraphics[width=0.9\linewidth]{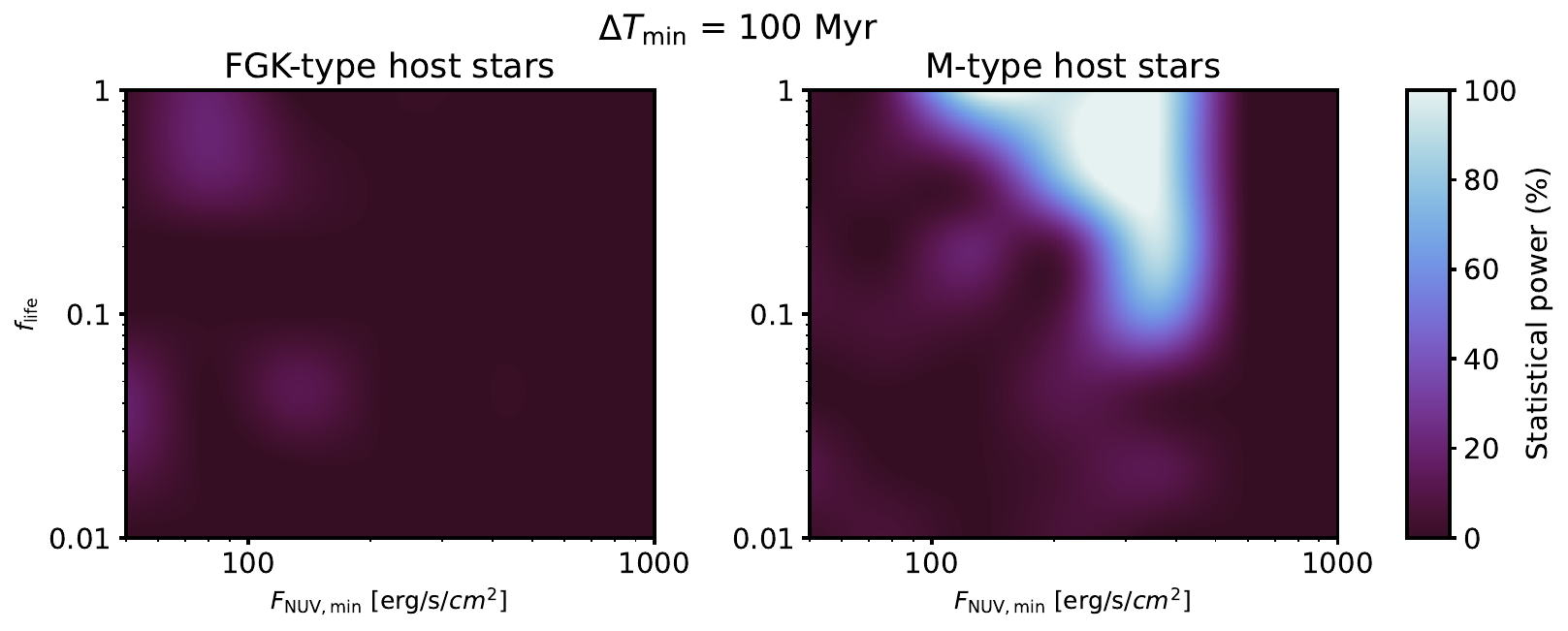}
    \caption{\rev{Statistical power for different abiogenesis rates and NUV threshold fluxes under the assumption of $\Delta T_{\mathrm{min}} = 100$ Myr. The longer origins timescale makes a test of the UV Threshold Hypothesis impossible for FGK planets, while M dwarf planets remain viable targets.}}
    \label{fig:powergrid_dT100}
\end{figure*}

\rev{
To assess the implications of a more conservative abiogenesis timescale, we consider a scenario in which a planet's minimum time required to be both in the \gls{HZ} and having above-threshold \gls{NUV} fluxes before life can emerge is increased to $\Delta T_{\mathrm{min}} = 100$ Myr, rather than 1 Myr as assumed above.
Figures~\ref{fig:inhabited_dT100}, \ref{fig:surveys_dT100}, and \ref{fig:powergrid_dT100} show the results of our \bioverse\ simulations under this assumption.
}

\rev{
The overall impact differs significantly between M dwarf and FGK dwarf planets.
 While M dwarf planets remain largely unaffected, the fraction of inhabited FGK dwarf planets decreases sharply due to the limited overlap between the time of high UV flux and habitable conditions (compare Figure~\ref{fig:hz_nuv_evo}).
Even under the optimistic assumption that the probability of abiogenesis is unity ($ f_{\mathrm{life}} = 1 $), the fraction of inhabited planets under the UV Threshold Hypothesis declines rapidly with increasing UV flux thresholds (Figure~\ref{fig:inhabited_dT100}).
}

\rev{
The impact of a longer origins timescale on testing the UV Threshold Hypothesis becomes evident when we repeat our example survey ($f_{\mathrm{life}} = 0.8$, $F_{\mathrm{NUV},\mathrm{min}} = \SI{300}{\erg\per\second\per\centi\meter\squared}$) under this assumption (Figure~\ref{fig:surveys_dT100}):
While the survey of M dwarf planets remains largely unaffected, the fraction of inhabited FGK planets is reduced to zero, making a conclusive test of the UV Threshold Hypothesis impossible.
}
\rev{
Figure \ref{fig:powergrid_dT100} demonstrates that this result is independent of the threshold \gls{NUV} flux and the intrinsic abiogenesis rate.
Given the lack of constraints on the timescale of abiogenesis, these findings highlight the advantages of focusing on M dwarf planets when searching for biosignatures.
If life emerges slowly, FGK planets may rarely, if ever, reach the inhabited stage, whereas M dwarf planets remain viable targets for future surveys.
If biosignatures were nevertheless detected on FGK planets despite the constraints imposed by a longer abiogenesis timescale, it would suggest that either the UV Threshold Hypothesis, as formulated here, is incorrect or that the timescale of abiogenesis is shorter than $\mathcal{O}(\SI{100}{\mega\year})$.
}

\bibliography{bib,coauthors}

\end{document}